\documentclass[namedreferences]{solarphysics}
\usepackage[hyperref,optionalrh,solaromanenum]{spr-sola-addons} % For Solar Physics
\usepackage{graphicx}                    % For eps figures, newer & more powerfull
\graphicspath{ {figures/} }
\usepackage{rotating}
\usepackage{color}                       % For color text: \color command
\usepackage{breakurl}                         % For breaking URLs easily trough lines
\usepackage{abbreviations}

\begin{document}

\begin{article}

\begin{opening}

\title{Interfacing MHD Single Fluid and Kinetic Exospheric Solar Wind Models and Comparing Their Energetics}
%\documentclass[11pt,a4paper]{article}
%\usepackage[T1]{fontenc}
%\usepackage[utf8]{inputenc}
%\usepackage{amsmath}
%\usepackage{amsfonts}
%\usepackage{amssymb}
%\usepackage{fullpage}
%\usepackage{graphicx}
%\usepackage{rotating}
%\usepackage{afterpage}
%\usepackage{xcolor}
%\usepackage{wrapfig}
%\usepackage{authblk}
%\usepackage{hyperref}
%\usepackage{natbib}
%\bibliographystyle{abbrvnat}
%\setcitestyle{authoryear,open={(},close={)}}
%\usepackage{abbreviations}
%\graphicspath{ {../../PhDthesis/chapters/SolarWind/image/} }
%

\author[addressref={aff1,aff2,aff3},corref,email={sofia.moschou@cfa.harvard.edu}]{\inits{S.P.}\fnm{Sofia-Paraskevi}~\lnm{Moschou}}%\sep
\author[addressref={aff2,aff4}]{\inits{V.}\fnm{Viviane}~\lnm{Pierrard}}%\sep ,email={viviane.pierrard@oma.be}
\author[addressref=aff1]{\inits{R.}\fnm{Rony}~\lnm{Keppens}}%\sep ,email={rony.keppens@kuleuven.be}
\author[addressref=aff5]{\inits{J.}\fnm{Jens}~\lnm{Pomoell}}%\sep

%\author{\inits{}\fnm{}~\lnm{}\orcid{}}
%\author{P.~\surname{Author-a}$^{1}$\sep
%        E.~\surname{Author-b}$^{1}$\sep
%        M.~\surname{Author-c}$^{2}$
%       }
%   \institute{$^{1}$ First affiliation
%                     email: \url{e.mail-a} email: \url{e.mail-b}\\
%              $^{2}$ Second affiliation
%                     email: \url{e.mail-c} \\
%             }
\address[id=aff1]{Centre for mathematical Plasma Astrophysics, Department of Mathematics, KU Leuven, Celestijnenlaan 200B, 3001 Heverlee, Belgium}
\address[id=aff2]{Royal Belgian Institute for Space Aeronomy, 3 Avenue Circulaire, B-1180 Brussels, Belgium}
\address[id=aff3]{Harvard-Smithsonian Center for Astrophysics, 60 Garden Street, Cambridge MA 02138, USA}
\address[id=aff4]{Universit\'{e} Catholique de Louvain, B-1348 Louvain-La-Neuve, Belgium}
\address[id=aff5]{Department of Physics, University of Helsinki, Gustaf H{\"a}llstr{\"o}min katu 2a, 00560 Helsinki, Finland}

\runningauthor{S.P. Moschou \textit{et al.}}
\runningtitle{Interfacing MHD and Kinetic Solar Wind Models}

\begin{abstract}

An exospheric kinetic solar wind model is interfaced with an observation-driven single fluid magnetohydrodynamic (MHD) model. Initially, a photospheric magnetogram serves as observational input in the fluid approach to extrapolate the heliospheric magnetic field. Then semi-empirical coronal models are used for estimating the plasma characteristics up to a heliocentric distance of 0.1AU. From there on a full MHD model which computes the three-dimensional time-dependent evolution of the solar wind macroscopic variables up to the orbit of the Earth is used. After interfacing the density and velocity at the inner MHD boundary, we compare with the results of a kinetic exospheric solar wind model based on the assumption of Maxwell and Kappa velocity distribution functions for protons and electrons respectively, as well as with \textit{in situ} observations at 1AU. This provides insight on more physically detailed processes, such as coronal heating and solar wind acceleration, that naturally arise by inclusion of suprathermal electrons in the model. We are interested in the profile of the solar wind speed and density at 1AU, in characterizing the slow and fast source regions of the wind and in comparing MHD with exospheric models in similar conditions. We calculate the energetics of both models from low to high heliocentric distances.

\end{abstract}

\end{opening}

%%%%%%%%%%%%%%%% write a paper related introduction %%%%%%%%%%%%%
\section{Introduction}
%The solar wind is the background within which all the space weather phenomena evolve and the close to Sun physical quantities describing it can serve as initial conditions in computational models that will examine their evolution as they propagate in the interplanetary space before they reach the Earth. Revealing the underlying physical processes, such as the heating and the acceleration, will be useful in predicting dangerous incidents and protecting terrestrial life and technological applications. 
Solar wind heating and acceleration mechanisms are still subjects of active research. In computational models, physical quantities estimated or observationally inferred close to the Sun serve as boundary or initial conditions that will examine the solar wind evolution and its underlying physics as it propagates through interplanetary space.
The solar wind plasma can be studied macroscopically through the magnetohydrodynamic (MHD) approach or microscopically when using the kinetic approach. In the following Subsections \ref{sub:MHDintro} and \ref{sub:kinintro}, we briefly review the main aspects of both approaches used in this paper, which are here for the first time interfaced in a global model. 

%%%%%%%%%%%%%%%%%%%% MHD %%%%%%%%%%%%%%%%%%%%%%%%%%%%

\subsection{MHD Heliospheric Modeling}\label{sub:MHDintro}

\cite{Wang90} presented an empirical relation between the expansion of magnetic flux tubes and the solar wind speed, showing that they evolve inversely proportional to each other. This assumption was tested using more than two decades of observations, and can give predictions of the solar wind speed at Earth. The model involves synoptic magnetograms of photospheric field, which allow a Potential Field Source Surface (PFSS, with the source surface typically at 2.5$\mathrm{R}_\odot$) extrapolation which quantifies the expansion factors. It was found that at the Earth's orbit, greater expansion corresponded to magnetic field lines near the centre of coronal holes, which diverge more slowly than the ones coming from the hole boundaries. This is consistent with the fact that lower densities are found in the fast wind regions.
\cite{Arge00} improved the Wang-Sheeley model by using daily updated magnetogram data from the Wilcox Solar Observatory (WSO) and relating the magnetic flux tube expansion factor with the solar wind speed at the source surface, while including effects of stream interactions from the source surface to the Earth. 
A statistical study which covered three years and compared the Wang-Sheeley model predictions with data from the \textit{Wind}\footnote{NASA spacecraft at the L$_1$ Lagrangian point of the Earth designed for long-term solar wind measurements and its effects on the terrestrial magnetosphere (\url{https://wind.nasa.gov}).} satellite was presented. The interplanetary magnetic field (IMF) polarity was properly predicted 75\% of the time, while solar wind speeds were within 10-15\% of actual values, when a 6-month period with data gaps was removed.

%%%%%%%%%%%%%%%%%%%% Euhforia %%%%%%%%%%%%%%%%%%%%%%%%%%%%
In the computational work of~\cite{Odstrcil99}, solar wind variations were examined in the corotating frame with a three-dimensional MHD model with a CME (Coronal Mass Ejection) injection scheme in the streamer belt.
Such MHD models take into account magnetic field variations due to the CME
interaction with the solar wind during the CME's evolution.
The CME movement depends on the background solar wind density and velocity and the vector properties of the solar wind magnetic field and velocity are affected by the passing disturbance. ENLIL~\citep{ENLIL1999a,ENLIL1999b,ENLIL2003} is a heliospheric MHD model that provides a three-dimensional description of the time-dependent solar wind evolution.  
It can use the Wang-Sheeley-Arge (WSA)~\citep{Wang90,Arge00} semi-empirical model as its boundary condition. 
The WSA model was used by~\cite{McGregor11} to study the solar wind at low heliocentric distances including an empirical method to link the magnetic field information with the velocity at 21.5$\mathrm{R}_\odot$. The new method was cross-validated using the 3D MHD code ENLIL and by comparing the results with observations at 1AU and at further distances as provided by \textit{Ulysses}. The estimation of the solar wind speed at 21.5$\mathrm{R}_\odot$ was indeed better than previous models and it captured both fast and slow solar wind.

Similar to ENLIL, we use a fully 3D MHD code EUHFORIA (European heliospheric forecasting information asset) that from 0.1AU onwards models the evolution of the plasma environment in the inner heliosphere. The code details are discussed in~\cite{euhforia} and in this paper we adopt it to get the macroscopic description of the solar wind.

%%%%%%%%%%%%%%%%%%%% Kinetic %%%%%%%%%%%%%%%%%%%%%%%%%%%%
\subsection{Kinetic Exospheric Models}\label{sub:kinintro}

Exospheric kinetic models are simplified collisionless, stationary models, that are meant to 
explain the acceleration of the solar wind in a self-consistent way and they were
first established by \cite{Jockers70} and \cite{Lemaire}.
The model was one-dimensional and time-independent
and provided the state of the solar wind plasma along a magnetic field line.
The acceleration of
the solar wind was due to the induced electric field even without suprathermal electrons
(Maxwellian distribution), but when accounting for the presence of suprathermal electrons, the terminal speed at 1AU increased.

The original exospheric model \citep{Jockers70, Lemaire} assumed Maxwellian velocity distribution functions (VDFs) for protons and electrons and supersonic winds of 300 $\mathrm{km\ s^{-1}}$ could be reached at 1AU with temperatures of the order of 1 MK for both species at the exobase (the distance beyond which collisions become negligible).   
Nevertheless, it remained difficult for the model to achieve higher bulk velocities, such as the ones observed in the fast solar wind, without increasing the temperature to unrealistic high values (10MK) at the exobase, or by adding other sources of solar wind acceleration. After the induced electric field is calculated, the solar wind acceleration spontaneously follows giving the solution of the solar wind from sub- to supersonic, without any extra energy terms assumed.

A Lorentzian (Kappa) velocity distribution function is used instead of the
classic Maxwellian in search for better agreement with observations
in~\cite{Pierrard96}. Indeed, suprathermal electrons are generally observed in
the velocity distribution functions measured \textit{in situ} in the solar wind.
\cite{Pierrard96} have shown that the presence of such suprathermal electrons
accelerates the wind to higher bulk velocities, so that no other source of
energy needs to be considered to reach the values observed in the high speed
solar wind.

\cite{Maks97} applied the kinetic model developed by~\cite{Pierrard96} using Kappa VDFs for both electron and proton populations that escape from the Sun to describe the solar wind. Since the first exospheric model, the semi-analytic kinetic model has been able to describe not only the fast but also the slow solar wind together with their sources, in the cold coronal hole and hot equatorial regions respectively, without unrealistic assumptions of too high temperatures and extra heating in the corona, as required by non-turbulence driven fluid models. 

While previous exospheric models placed the exobase at a distance of about
5-10$\mathrm{R}_\odot$, from where on the proton total potential energy was a
monotonic function of the heliocentric distance, \cite{Lamy03} calculated the
exobase to be positioned at about 1.1-5$\mathrm{R}_\odot$. This deeper location of the exobase, lowered under the radial location of the maximum of the total potential energy of the protons, gives the solar wind the observed acceleration to high velocities. A low exobase leads indeed to higher bulk velocities at 1AU in the case of suprathermal electrons.

Collisionless (exospheric) theoretical models and collisional simulations were compared in \cite{Zouganelis05}. Including suprathermal tails in the velocity distribution function of the electrons and employing a self-consistently computed heat flux,
the models were able to reproduce fast solar wind speeds. Results of collisional kinetic
simulations with non-Maxwellian velocity distribution functions and
collisionless exospheric models are in good agreement. Taking into account that the exospheric and collisional models provide comparable results, in this paper we will go a step further and try to interface exospheric models with MHD ones. 

On the way to developing predictive tools and 3D solar wind models, a 2D
observationally driven kinetic exospheric solar wind model was developed
by~\cite{Pierrard14pieters}, presenting solar wind variations on the ecliptic
plane and how they compare to observations from close to the Sun up to 1AU.
For the ecliptic variational study OMNI\footnote{Multi-source data set for the near Earth solar wind of combined and normalized observational data from ACE (\textit{Advanced Composition Explorer}), \textit{Wind}, IMP 8 (\textit{Interplanetary Monitoring Platform}) and GOES (\textit{Geostationary Operational Environmental Satellite}) satellite missions.} observations were used for the time period 26 September to 23 October 2008.
% An exobase at 1.1$\mathrm{R}_\odot$ was assumedand particle temperatures of $T_e=1$MK and $T_p=2$MK, respectively, to examine the latitudinal dependence of the exospheric model in the range $[-45^\circ,45^\circ]$. \comm{this range does not mean anything by itself, it depends completely on if there are coronal holes etc.}
The $\kappa$ parameter was chosen as 2.35 and 3.82 for
fast and slow wind respectively, to match bulk speed observations close to the
orbit of the Earth. We will present a three dimensional generalization of this
exospheric model, \textit{i.e.}\ we find the solar wind characteristics along a
collection of magnetic field lines each passing through a point on the spherical
shell at the exobase level in latitude and longitude $(\theta,\phi)$.

The basic principles, boundary conditions, physical assumptions and
computational methods used by the MHD and the exospheric kinetic models are
described in Section 2. The specific criteria and the observational data that
are chosen for this work are explained and presented in Section 3.
The interfacing method as well as explicit results of both approaches and their energetics are discussed in Section 4, while in Section 5 we compare the two approaches and we close by discussing the main conclusions of the study in Section 6.

\section{Models}
%%%%%%%%%%%%%%%%%%%%%%%%%%%%%%%%%%%%%%%%%%%%%%%%%%%%%%%%%%%%%%%%%
%%%%%%%%%%%%%%%%%%%%%%%%%%%% MHD %%%%%%%%%%%%%%%%%%%%%%%%%%%%%%%%%%
\subsection{MHD Modeling: EUHFORIA}
%%%%%%%%%%%%%%%%%%%%%%%%%%%%%%%%%%%%%%%%%%%%%%%%%%%%%%%%%%%%%%%%%
% \subsubsection{Euhforia project outline}

The inner heliosphere model EUHFORIA~\citep{euhforia} is used for our MHD
approach. EUHFORIA is a three-dimensional observationally driven model
providing an accurate description of the large-scale
time-dependent solar wind including transient events such as CMEs.
As such, it allows to inject CMEs at the inner radial boundary
at 0.1AU as a time-dependent boundary condition.
Apart from CMEs, the variables at the inner radial boundary
are constructed in order to capture the large-scale
variations in the solar wind for the particular time period under study.
This is accomplished using a model for the coronal magnetic field and employing
empirical relations between the coronal magnetic topology and the state
of the solar wind. The magnetic field model
consists of a potential field source surface (PFSS) model in the low corona
coupled with a current sheet model higher in the corona. The PFSS
model requires a magnetogram to be provided as input.
To finally compute the super-sonic state of the solar wind at 0.1AU,
empirical relations inspired by the success of
the WSA (Wang-Sheeley-Arge) model~\citep{Wang90,Arge00}
are used.

The MHD model is able to provide density and speed profiles at the Earth's
orbit, it allows for slow and fast solar wind source region tracing, and can
serve as the MHD counterpart in a comparison project together with kinetic
exospheric models that correspond to similar initial and boundary conditions at
0.1AU. This will be our first goal in this paper, and the way the two
approaches are coupled will be described next.

 %%%%%%%%%%%%%%%%%%%%%%%%%%%%%%%%%%%%%%%%%%%%%%%%%%%%%%%%%%%%%%
 \subsubsection{MHD Equations, Methods and Schemes}

EUHFORIA uses a finite volume discretization scheme to solve the hyperbolic conservative MHD equations.
The equations solved are those of ideal MHD with gravity included as a source term in the equations of momentum and energy:
\begin{eqnarray}
\label{eq:rho}
\frac{\partial \rho}{\partial t}+\nabla\cdot(\rho \mathbfit{v})=0{,} \\
\label{eq:forces}
\frac{\partial(\rho\mathbfit{v})}{\partial t}+\nabla\cdot\left[ \rho\mathbfit{v}\mathbfit{v}+\left( p+\frac{B^2}{2\mu_0}\right)\mathcal{I}-\frac{1}{\mu_0}\mathbfit{B}\mathbfit{B}\right]=\rho \mathbfit{g}{,} \\
\label{eq:p}
\frac{\partial\mathcal{E}}{\partial t}+\nabla\cdot\left[\left(\mathcal{E}+p-\frac{B^2}{2\mu_0}\right)\mathbfit{v}+\frac{1}{\mu_0}\mathbfit{B}\times\left( \mathbfit{v}\times\mathbfit{B}\right)\right]=\rho\mathbfit{v}\cdot\mathbfit{g}{,} \\
\label{eq:B}
\frac{\partial \mathbfit{B}}{\partial t}-\nabla\times\left( \mathbfit{v}\times\mathbfit{B}\right)=0{,} \\
\label{eq:sol}
\nabla\cdot\mathbfit{B}=0{,} \\
\label{eq:energy}
\mathcal{E}= \frac{p}{\gamma-1}+\frac{\rho v^2}{2}+\frac{B^2}{2\mu_0} {,} \qquad \gamma=1.5
\end{eqnarray}
where $\rho$ is the mass density,
$\mathbfit{v}$ the velocity vector,
$\mathbfit{g}$ the gravitational acceleration,
$\mathbfit{B}$ the magnetic field vector,
$p$ the thermal pressure,
$\gamma$ the polytropic index,
%$\mathbfit{E}$: the electric field vector,
%$\mathbfit{j}$: the current,
$\mathcal{E}$ the total energy density,
$\mu_0$ the magnetic permeability and
$\mathcal{I}$ the unit tensor.
Note that we are working in the inertial frame, so no Coriolis nor centrifugal forces need to be added in Equation \ref{eq:forces}.
The polytropic index is chosen to be slightly smaller than the expected
$\gamma=5/3$ value for a monatomic gas. This causes a
finite energy to be injected into the system in the
form of heat \citep[see \textit{e.g.}][and references therein]{Pomoell2012}.
The use of either a non-adiabatic polytropic index or explicit source terms in
the momentum and energy equation to drive the solar wind and heat the corona have been
used in several works. In EUHFORIA, the reduced polytropic index is used in order to slightly accelerate the solar wind further out in the heliosphere, from the speed values at the boundary at 0.1AU. The single fluid MHD description still leaves freedom to vary $\gamma$, which allows to account for expected deviations from the mono-atomic ideal gas value of 5/3. 
For a discussion of more self-consistent models  that attempt to capture and explain the physical mechanisms resulting in the observed coronal heating and acceleration, we refer to~\cite{Cranmer2012}.

The employed numerical grid is uniform in spherical coordinates,
with the number of cells in $r$, $\theta$, $\phi$ chosen to be
$800$, $60$, $180$, respectively. The outer boundary is set at 2AU. 
Further details of the numerical solution scheme are described in \cite{euhforia}.

 %%%%%%%%%%%%%%%%%%%%%%%%%%%%%%%%%%%%%%%%%%%%%%%%%%%%%%%%%%%%%%
\subsubsection{Boundary Conditions}

The essential input to EUHFORIA is a synoptic magnetogram.
We select a magnetogram from the GONG (Global Oscillation Network Group) standard synoptic data product, which are available with one hour cadence from GONG. The chosen magnetogram corresponds closely to Carrington Rotation (CR) 2059.
During this Carrington rotation, an equatorial coronal hole was visible near the central meridian to about $60^\circ$ degrees west. The solar wind plasma state at 0.1AU in EUHFORIA is determined using a
semi-empirical approach similar to the Wang-Sheeley-Arge model. The method
consists of constructing a model of the coronal magnetic field consisting
of a PFSS extrapolation in the low corona
while the "Schatten" Current Sheet is used from
$2.5\mathrm{R}_\odot$ to 0.1AU. The solar wind speed is then given through an empirical relation which is a function of the magnetic flux tube expansion factor. The formula used in this work is
given by
\begin{eqnarray}
\label{eq:expansionv}
\label{eq:empjens}
V(f_s)=240.0+675.0(1+f_s)^{-0.22} \mathrm{km\ s^{-1}} {,}
\end{eqnarray}
where $f_s$ is given by
\begin{eqnarray}
\label{eq:fs}
f_s=\left(\frac{\mathrm{R}_\odot}{r}\right)^2\frac{B_r(\mathrm{R}_\odot,\theta_0,\phi_0)}{B_r(r,\theta,\phi)}
\end{eqnarray}
and it quantifies the expansion factor of the flux tube from the photospheric footpoint $(\mathrm{R}_\odot,\theta_0,\phi_0)$ of the specific field line to its position further outwards $(r,\theta,\phi)$ at a heliocentric distance $r$~\citep{Wang97}. As explained in \cite{Wang97}, the expansion factor takes values greater or equal to unity for flux divergence more rapid than or equal to $r^2$, respectively.
Simple scaling laws that are functions of $V$ are used in order to
determine the plasma density and temperature.
For further details of the empirical model, see \cite{euhforia}.

%%%%%%%%%%%%%%%%%%%%%%%%%%%%%%%%%%%%%%%%%%%%%%%%%%%%%%%%%%%%%%
%%%%%%%%%%%%%%%%%%%%%%%%%%%%%%%%%%%%%%%%%%%%%%%%%%%%%%%%%%%%%%%%%
%%%%%%%%%%%%%%%%%%%%%%%%%%% KINETIC %%%%%%%%%%%%%%%%%%%%%%%%%%%%%%%%%
%%%%%%%%%%%%%%%%%%%%%%%%%%%%%%%%%%%%%%%%%%%%%%%%%%%%%%%%%%%%%%%%%

\subsection{Kinetic Exospheric Model}

For the kinetic component of our analysis, we are using an exospheric model,
which is a way to simulate low density plasmas, where the importance
of collisions is limited. 
The solar atmosphere is considered to have a
collision-dominated barosphere at low altitude~\citep[below 1.1\,--\,10$\mathrm{R}_\odot$ according to][]{Lamy03} and a collisionless exosphere,
which is the region modeled kinetically. These regions are separated by a
surface called the exobase $r_0$, beyond which collisions become negligible. This
exobase level is defined as the altitude where the particle mean free path
$l_f$ and the local density scale height $H$ become equal, \textit{i.e.} where the
dimensionless Knudsen number $K_n=l_\mathrm{f}/H$ is equal to unity. The kinetic model\footnote{A 1D version of the kinetic exospheric model
developed by the group in IASB-BIRA and collaborators can be found in CCMC
(\url{http://ccmc.gsfc.nasa.gov/models/exo.php}) and it can run online for
user-defined setups.}~\citep{Lemaire,Pierrard96,Maks97,Lamy03} gives different
temperatures for electrons and protons as indeed observed \citep{Lemaire01} and
can include different characteristics of any other ion species.

\begin{figure}[htbp]
\begin{center}
\includegraphics[width=0.8\textwidth]{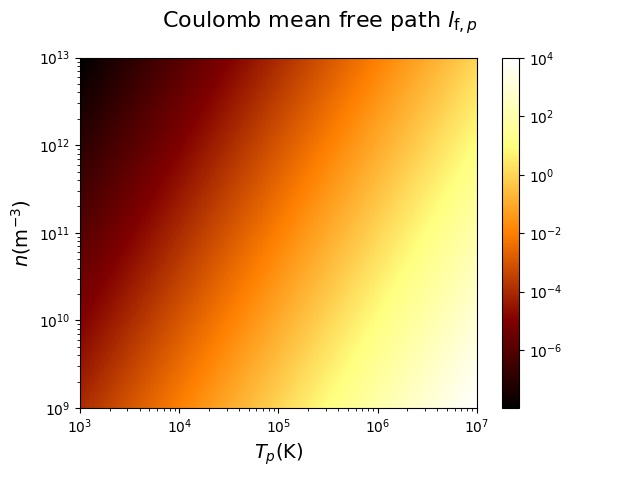}
\caption{The Coulomb mean free path $l_{\mathrm{f},p}$ in solar radii as a function of proton temperature and electron density. This figure quantifies the variations in Equation \ref{eq:kn}.}
\label{fig:lam}
\end{center}
\end{figure}

Sources of the fast solar wind are considered to be coronal holes and in
these regions the electron VDFs are assumed to correspond to a Lorentzian function with a small
$\kappa$-value and thus have a large suprathermal tail~\citep{Maks97}. The low speed solar wind usually comes from equatorial regions, with larger $\kappa$-values.
When $\kappa \rightarrow \infty$ the VDF tends to a Maxwellian.
In this work, we consider two particle species, namely electrons and protons and
therefore their respective exobase levels need to be defined. The proton exobase is located
where the Coulomb mean free path for the protons $l_{\mathrm{f},p}$ according
to~\cite{Spitzer62} as estimated for coronal values by~\cite{Maks97} is equal to
the local density scale height $H$, where
\begin{eqnarray}
l_{\mathrm{f},p}\approx7.2\times10^7\frac{T^2_p}{n_e}{,} { \ \ \ \  }
H=\left(-\frac{\mathrm{d}\ln n_e}{\mathrm{d}r} \right)^{-1} {,}
\label{eq:kn}
\end{eqnarray}
where all the quantities are in SI. The proton mean free path is shown in
Figure~\ref{fig:lam} as a function of the proton temperature and the electron
number density. For the electrons, a similar electron exobase height can be
estimated from the Coulomb mean free path in a plasma consisting
only of electrons and protons:
\begin{eqnarray}
l_{\mathrm{f},e}=0.416 \left(\frac{T_e}{T_p}\right)^2l_{\mathrm{f},p} {,}
\end{eqnarray}
as in~\cite{Maks97} for a hydrogen plasma, assuming that the
electrons have the mean thermal velocity $(8k_\mathrm{B}T_e/m_e\pi)^{1/2}$, with
$l_{\mathrm{f},e},l_{\mathrm{f},p}$ in meters and $T_e,T_p$ in kelvin. A crude estimate of the scale height can be obtained assuming hydrostatic equilibrium
in a stratified atmosphere with isothermal plane parallel layers.
This is not the case for the solar wind, since expansion is taking place and
rather hydrodynamic conditions apply, but it gives a good approximation of the
order of magnitude of the scale height.

When we adopt the same proton and electron temperature in the above formulae, the electron mean free path becomes smaller than the proton one
$l_{\mathrm{f},e}<l_{\mathrm{f},p}$, such that the electron collisions are more important for higher
altitudes and thus the proton exobase is found at lower
altitudes~\citep{Maks97}. In this study, we make the assumption that both
populations have the same exobase altitude, and we choose it to correspond to
the source surface location $r_{0,p}=r_{0,e}=r_0=2.5\mathrm{R}_\odot$ where we by construction obtain purely 
radial magnetic fields. The comparison between $l_\mathrm{f}$ and $H$
shows anyway that the collisions become negligible already at very low radial
distances in the solar corona. Some indicative values for the different source
regions on the Sun, namely coronal hole and equatorial regions, are estimated by~\cite{Hundhausen68} and \cite{Withbroe88}.
Figure~\ref{fig:lam} illustrates the Coulomb mean free path $l_{\mathrm{f},p}$ as a
function of temperature and number density to show the possible positions of the exobase. According to~\cite{Lamy03}, the exobase for equatorial regions is estimated to be at about 5\,--\,10$\mathrm{R}_\odot$, whereas for coronal holes the exobase is estimated to be positioned at about 1.1\,--\,5$\mathrm{R}_\odot$. \cite{Scudder13} have shown that suprathermal particles are already
collisionless for $K_n>0.01$, due to the velocity dependence of the mean free path of
the particles. This shows that it is not especially important that the exobase
is chosen to correspond exactly to the level where $K_n=1$, but it will appear
where the density gradient is very sharp and thus where the plasma becomes
collisionless to a good approximation.

%%%%%%%%%%%%%%%%%%%%%%%%%%%%%%%%%%%%%%%%%%%%%%%%%%%%%%%%%%%%%%

\subsubsection{Velocity Distribution Functions}
%Basics of the Solar Wind p.66
When collisions are ignored as in the exospheric theory developed
by~\cite{Lemaire}, the Fokker-Planck equation reduces to the Vlasov equation for
the evolution of the velocity distribution function:
\begin{equation}
\frac{\partial f}{\partial t}+\mathbfit{v}\cdot\frac{\partial f}{\partial \mathbfit{r}}+\mathbfit{a}\cdot\frac{\partial f}{\partial \mathbfit{v}}=0 {.}
\end{equation}
Our kinetic exospheric model works by constructing stationary solution to the
Vlasov equation, starting from an exact stationary solution for protons and
electrons prescribed at the exobase. Kinetic models based on this equation were
developed and are discussed in~\cite{Maks97} for radial magnetic field lines and
in~\cite{Pierrard01meyer} taking into account the spiral interplanetary magnetic
field topology. It was shown in~\cite{Maks97} that the specific moments in the
solar wind, namely densities and temperatures, as well as the electrostatic
potential characteristics from the corona to the interplanetary space are
already well described, agreeing with observations at 1AU, when the collision
term is neglected, since the collisions would rather modify the temperature
anisotropies.

Apart from the Maxwellians the generalised Lorentzian or Kappa function is also
a solution of the Vlasov equation and can be used as a boundary condition to
study the effect of suprathermal particles on the kinetic moments. Observations
suggest that the velocity distribution functions of the electrons have strong
suprathermal tails. We therefore assume a Lorentzian VDF for the electrons and a
Maxwellian VDF for the protons at the exobase:
\begin{eqnarray}
\label{eq:fp}
& f^p_{\mathrm{Maxwell}}(r_0,v)=n_p(r_0)\left(\frac{m_p}{2\pi k_\mathrm{B}T_p(r_0)} \right)^{3/2}\exp\left( -\frac{m_pv_p^2}{2k_\mathrm{B}T_p(r_0)}\right) {,} \\
\label{eq:fe}
& f^e_{\mathrm{kappa}}(r_0,v)=\frac{n_e(r_0)}{2\pi\kappa^{3/2}}\left(\frac{m_e}{2k_\mathrm{B}T_e(r_0)} \right)^{3/2}A(\kappa)\left(1+\frac{m_ev_e^2}{2k_\mathrm{B}T_e(r_0)\kappa} \right)^{-(\kappa+1)} {,} \qquad
\end{eqnarray}
where
\begin{equation}
A(\kappa)=\frac{\Gamma(\kappa+1)}{\Gamma(\kappa-1/2)\Gamma(3/2)} {.}
\end{equation}
We note in passing that the moments of the Lorentzian VDF are not well defined for
every $\kappa$ value, but rather every $i$th moment is defined for
$\kappa>(i+1)/2$~\citep{Pierrard96}.
Suprathermal protons have almost no influence on the solar wind velocity, so for them a Maxwellian VDF can suffice~\citep{Maks97}.

%%%%%%%%%%%%%%%%%%%%%%%%%%%%%%%%%%%%%%%%%%%%%%%%%%%%%%%%%%%%%%
Liouville's theorem~\citep{goldstein2002classical} implies that any
function that depends on the constants of motion of a collection of particles
satisfies the Vlasov equation. The relevant constants of motion in this study
are the total energy and the magnetic moment. Knowing the velocity distribution
functions for our particle species at the exobase, the velocity distribution as
a function of the radial distance can be deduced from energy
conservation~\citep{Pierrard96}. Thereby, the electron and proton VDFs can be computed as a function of radius and
speed along a purely radial magnetic field line. The analytic expressions for the kinetic moments of the exospheric models were
derived for a Maxwellian VDF by~\cite{Lemaire} and for a Lorentzian VDF
by~\cite{Pierrard96}. As explained above the exospheric model used
in this paper includes only radial velocities along open radial magnetic field lines.

Like in previous exospheric models, it is assumed that there are no particles
coming from the interplanetary space to the Sun. The anisotropy of the
distribution leads to the solar wind flux. The density, temperature and
$\kappa$-index are determined at the exobase by either the MHD model or
constrained \textit{via} observations. The model provides then the velocity distribution function at any
other distance as well as the rest of the kinetic moments. We use the code and
the analytical expressions for the Maxwellian and the $\kappa$ distributions
by~\cite{Pierrard96}. Provided the number density, the electron and proton
temperatures and the $\kappa$ index for the electron VDF at the exobase, the
quasi-neutrality and zero-current conditions are solved iteratively at a fixed
radial distance $r_m$ using a Newton-Raphson scheme. The value of $r_m$ is
iteratively modified using a dichotomy method until the electric field is found
to be continuous within a predefined tolerance~\citep{Lamy03}.

On the other hand, EUHFORIA solves the 3D MHD equations
taking self-consistently stream interactions into account thereby
providing a $v_\phi$ for any point, whereas in the kinetic model we
impose that this velocity is constant on each spherical
shell. The kinetic model thus proceeds without accounting for stream interactions
instead keeping the same topology of fast and slow solar wind sources at every
radial distance as at the exobase, reducing the computational time. As was
argued in~\cite{Pierrard01meyer}, the effects due to rotation
as compared to the purely radial case
change only the estimated proton and electron temperatures
and their anisotropies. More specifically, the spiral structure predicts higher
electron temperatures $T_e$ and lower proton temperatures $T_p$ than the radial
case, but the number density, the electric potential and the bulk speed remain
almost unaffected up to $300\mathrm{R}_\odot$. It's worth noting here that stream interactions are once again not taken into account.
Therefore, in this study we use the radial
magnetic field topology and simply rotate by $v_\phi$ each spherical
shell to account for solar rotation.

The advantage of the kinetic model used in this paper is the
direct quantification of species-specific temperature profiles, densities,
speeds, energy fluxes \textit{etc}. once the electric potential is calculated. Even if $T_e=T_p$ is chosen at the exobase, the kinetic model self-consistently calculates the species-specific heating with distance and the two temperatures depart from each other. The $T_e$ is indeed observed to be different than $T_p$ at 1AU for slow and fast wind cases, \textit{e.g.} as reported in~\cite{Lemaire01}.

%%%%%%%%%%%%%%%%%%%%%%%%%%%%%%%%%%%%%%%%%%%%%%%%%%%%%%%%%%%%%%%%%
%%%%%%%%%%%%%%%%%%%%%%%%%%%%%%%%%%%%%%%%%%%%%%%%%%%%%%%%%%%%%%%%%

\section{Observational Input: Cases and Selection Criteria}

Several missions have observed, or continue to observe the Sun, as well as
measure the physical parameters that characterize the solar wind, at different
heliocentric distances as well as heliographic latitudes. They provide high
resolution data, not only in the ecliptic plane, but also at higher latitudes,
requiring simulations that explain and predict the behavior of the solar wind
not only in the ecliptic plane, but rather in three dimensions. In this study we
will be using OMNI data, and data
from the \textit{Ulysses} spacecraft due to its large latitudinal coverage. We based our synoptic magnetogram and solar state selection on the
following criteria, that allow us to perform a crosscheck on their prediction
ability with available spacecraft observations:
i) quiet Sun periods (since the exospheric models are particularly tailored to quiet Sun conditions);
ii) the presence of equatorial coronal holes, such that significant differences between high and slow speed wind may be expected at the orbit of the Earth; iii) position of \textit{Ulysses} for combination of simulations and different spacecraft observations from different angles/telescopes; and iv) very few CME events according to the available catalogue CACTUS\footnote{More information can be found at \url{http://sidc.oma.be/cactus/}.}.

In this study we focus on the year 2007, as it was a mostly quiet Sun year coinciding with the third orbit of \textit{Ulysses}.
For the global 3D comparison we focus on the period July-August 2007, when \textit{Ulysses} crossed the ecliptic plane.
We have confirmed the relative paucity of CME events in the selected time period using the CACTUS CME list.

%%%%%%%%%%%%%%%%%%%%%%%%%%%%%%%%%%%%%%%%%%%%%%%%%%%%%%%%%%%%%%
\section{Analysis and Results}

%%%%%%%%%%%%%%%%%%%%%%%%%%%%%%%%%%%%%%%%%%%%%%%%%
\subsection{Constraining the Kinetic Model using EUHFORIA}

First, we constrain the input to the kinetic model based on data-driven results at the inner radial boundary of EUHFORIA at $21.5 \mathrm{R}_\odot$ and examine both models' efficiency at capturing the solar wind bulk quantities by comparing with observations at 1AU and at 1.4AU (the \textit{Ulysses} orbit). We compare the results of the MHD and kinetic models after interfacing their velocities and densities at the inner MHD boundary at $21.5\mathrm{R}_\odot$. 
 %%%%%%%%%%%%%%%%%%%%%%%%%%%%%%%%%%%%%%%%%%%%%%%%%%%%%%%%%%%%%%

\subsubsection{EUHFORIA}

 \begin{figure}[htbp]
\begin{center}
\includegraphics[width=\textwidth]{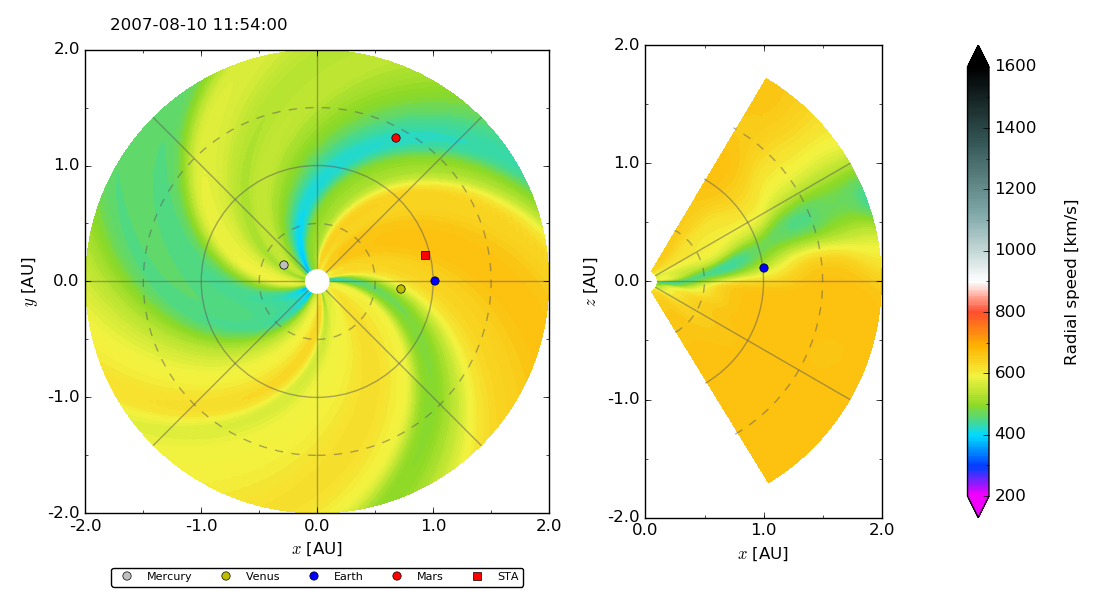}
\includegraphics[width=\textwidth]{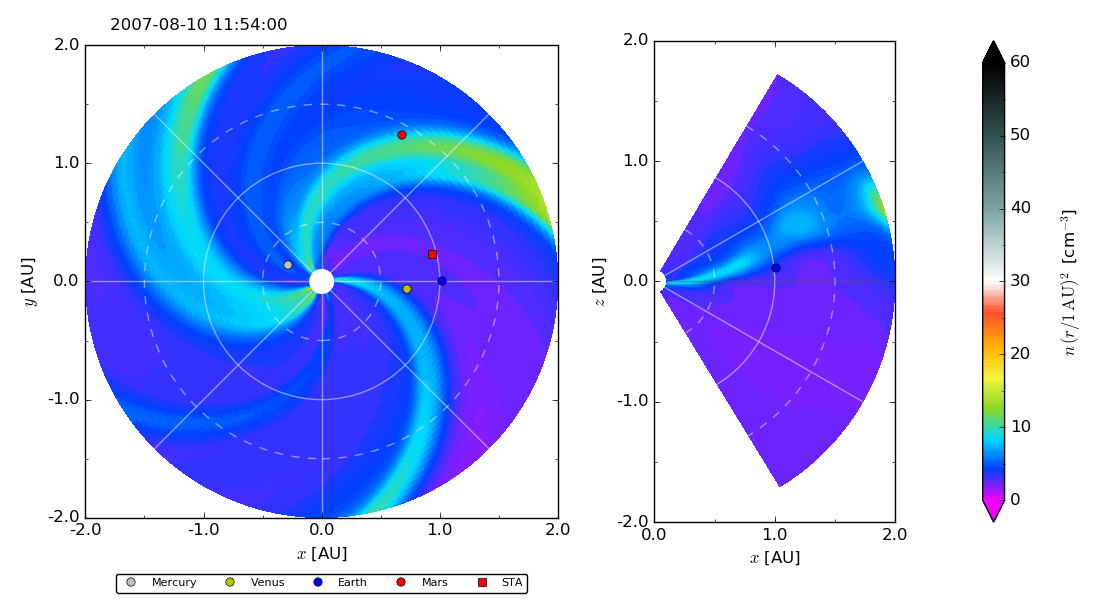}
\caption{EUHFORIA longitudinal and latitudinal variations of the radial velocity and the number density of the solar wind corresponding to the top and bottom rows, respectively for CR2059 of August 2007. \textit{Left panels}: on the equatorial plane and \textit{Right panels}: meridional plane including Earth.}
\label{fig:eufscaleaug}
\end{center}
\end{figure}

In Figure~\ref{fig:eufscaleaug} we show a slice at the equator (left panels) and
a slice in latitude (right panels) that corresponds to the mass density scaled with the inverse square of the
heliocenteric distance,
and the radial speed also indicating the positions of Mercury, Venus, Earth,
Mars and the STEREO (\textit{Solar Terrestrial Relations Observatory}) A spacecraft. The grey part of the colorscale
used in the Figure corresponds to high velocities and densities
that occur especially during CME events. The velocity ranges
ocurring at this time are roughly from 350 to 650 $\mathrm{km\ s^{-1}}$ with a
clear separation between streams of different speeds, forming a clear Parker
spiral-like configuration. There is a configuration of several distinct high and slow
speed streams. We can see the different streams with high density associated to
low speed and vice versa, while the highest speed captured is around $650
\mathrm{km\ s^{-1}}$ and corresponds to a density similar to the one measured
at 1AU. The highest density contrast with respect to the one measured at 1AU is
about 10. The latitudinal panels suggest that there is compression and
rarefaction as the plasma flows outwards with the corresponding speed and
density. We thus conclude that the plasma is not following exactly
the ideal Parker spiral moving on perfect cones as it expands, but following a
rather more complicated motion, as the simulation is observation-driven and the
photospheric magnetogram shows a complex topology, as discussed previously.

%%%%%%%%%%%%%%%%%%%%%%%%%%%%%%%%%%%%%%%%%%%%%%%%%

\begin{figure}[htbp]
\begin{center}
\includegraphics[trim={0mm 10mm 0mm 0mm},clip=true,width=0.51\textwidth]{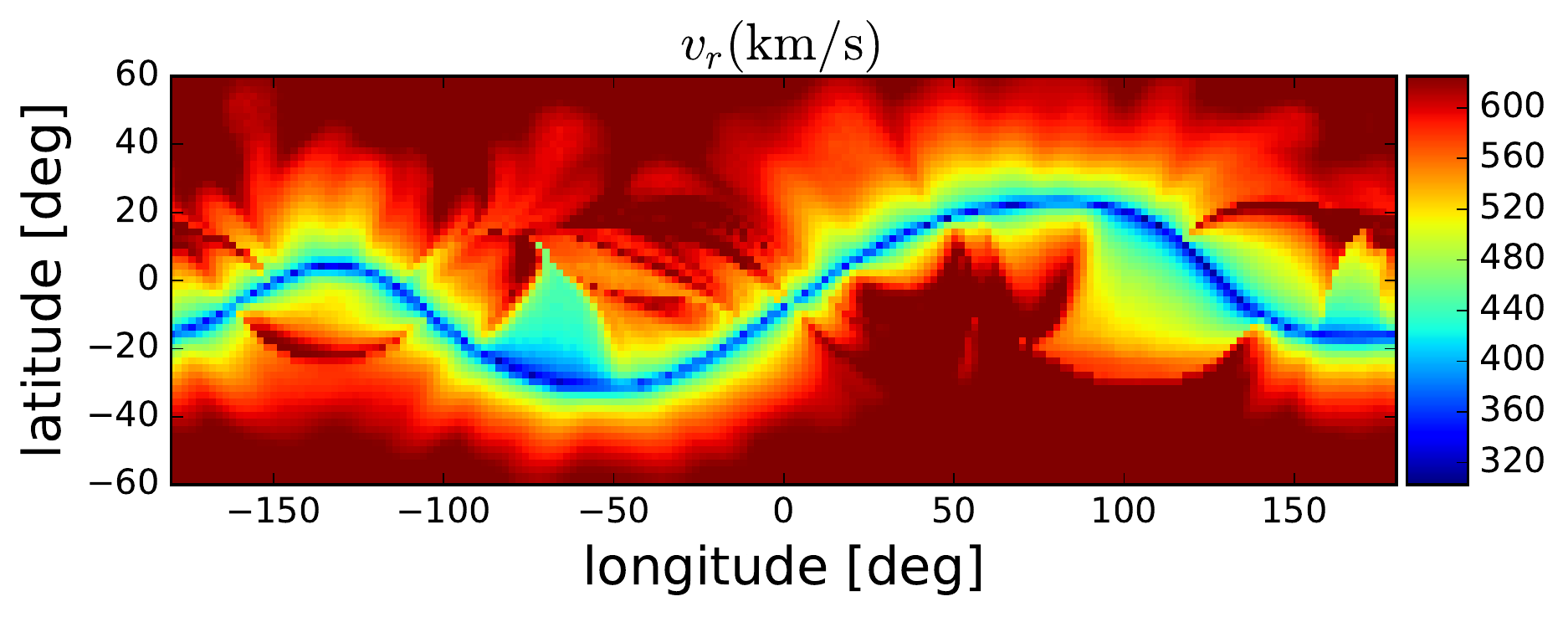}
\includegraphics[trim={10mm 10mm 0mm 0mm},clip=true,width=0.48\textwidth]{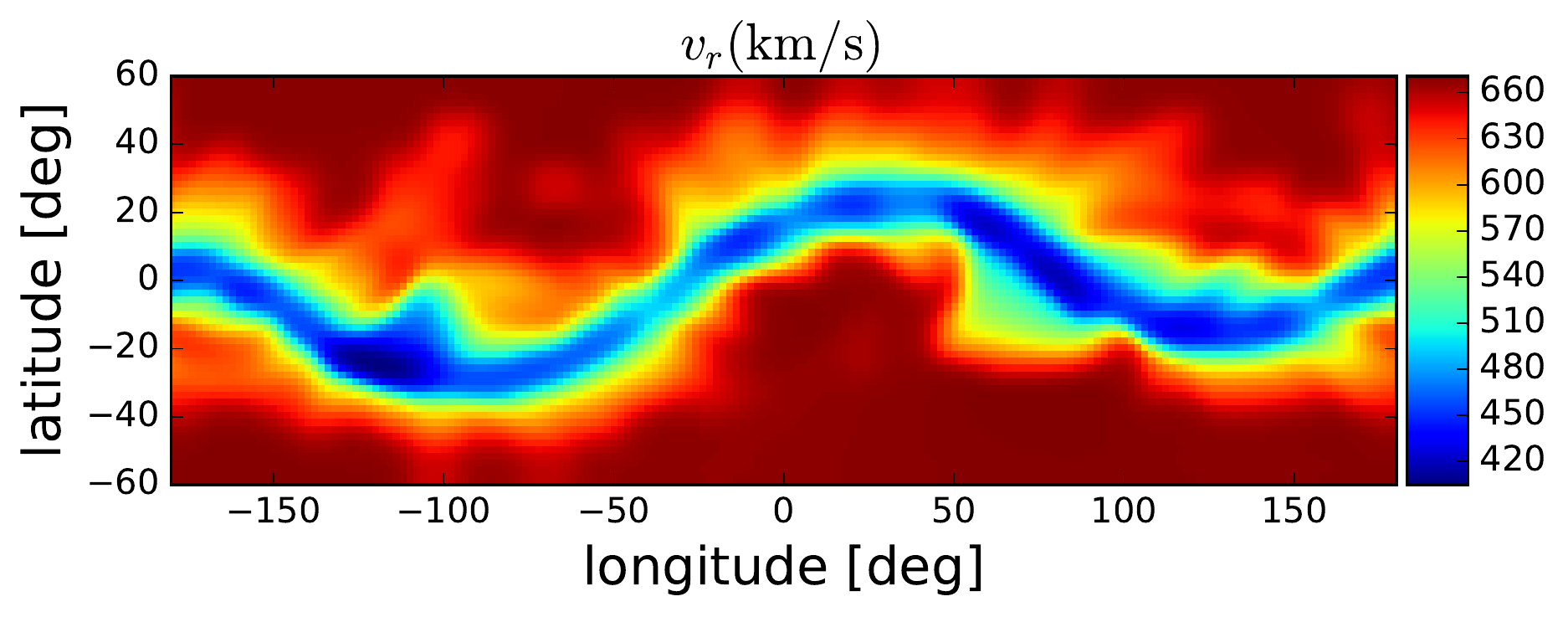}
\includegraphics[trim={0mm 10mm 0mm 0mm},clip=true,width=0.51\textwidth]{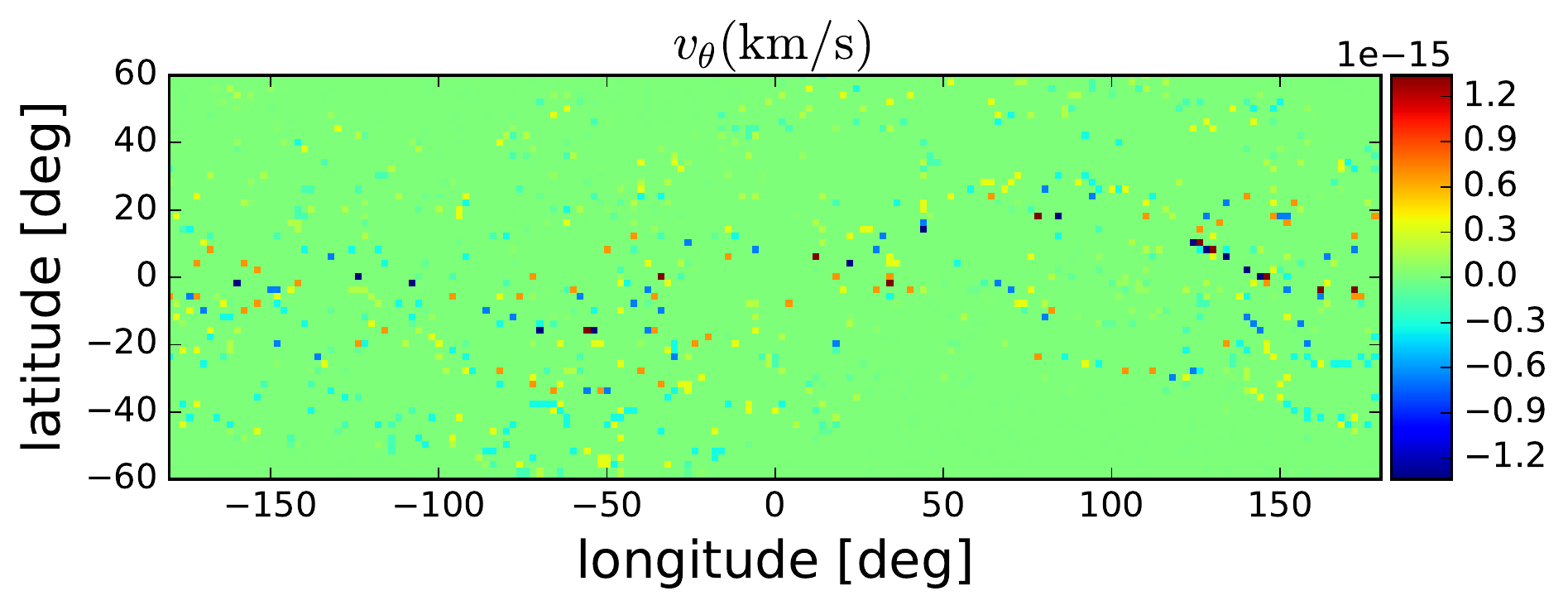}
\includegraphics[trim={10mm 10mm 0mm 0mm},clip=true,width=0.48\textwidth]{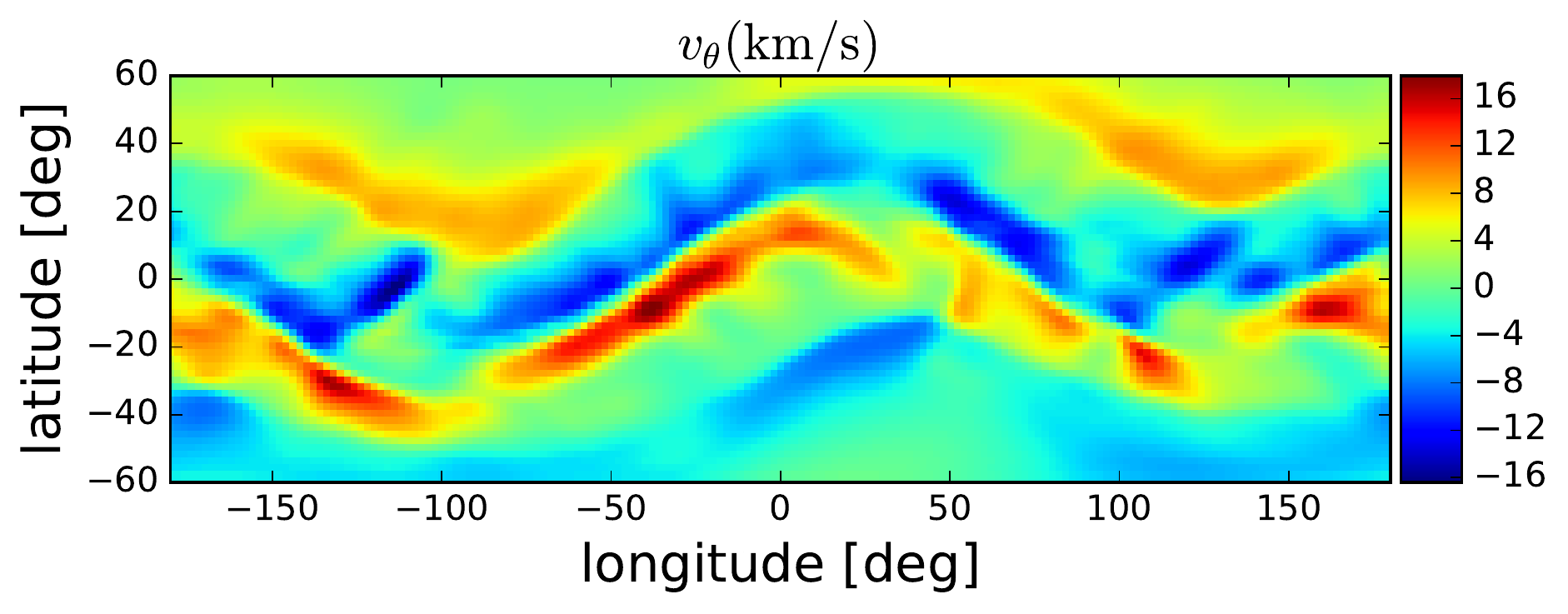}
\includegraphics[trim={0mm 10mm 0mm 0mm},clip=true,width=0.51\textwidth]{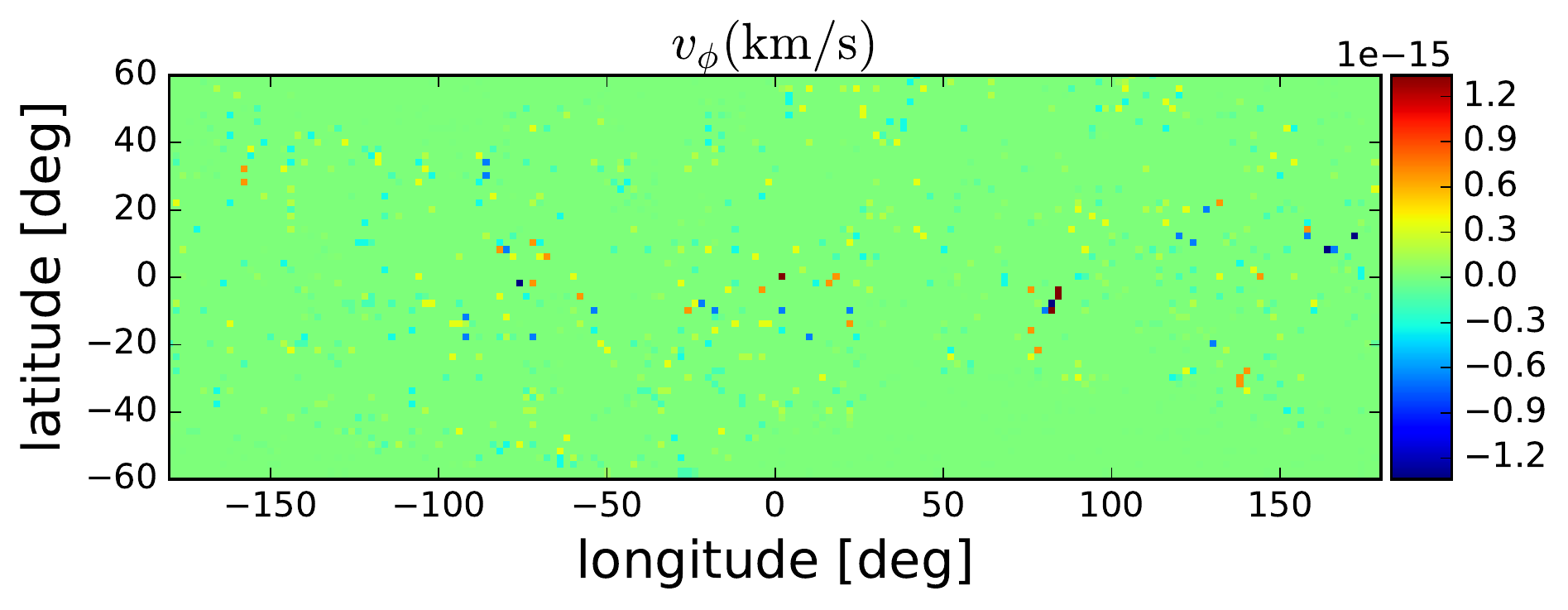}
\includegraphics[trim={10mm 10mm 0mm 0mm},clip=true,width=0.48\textwidth]{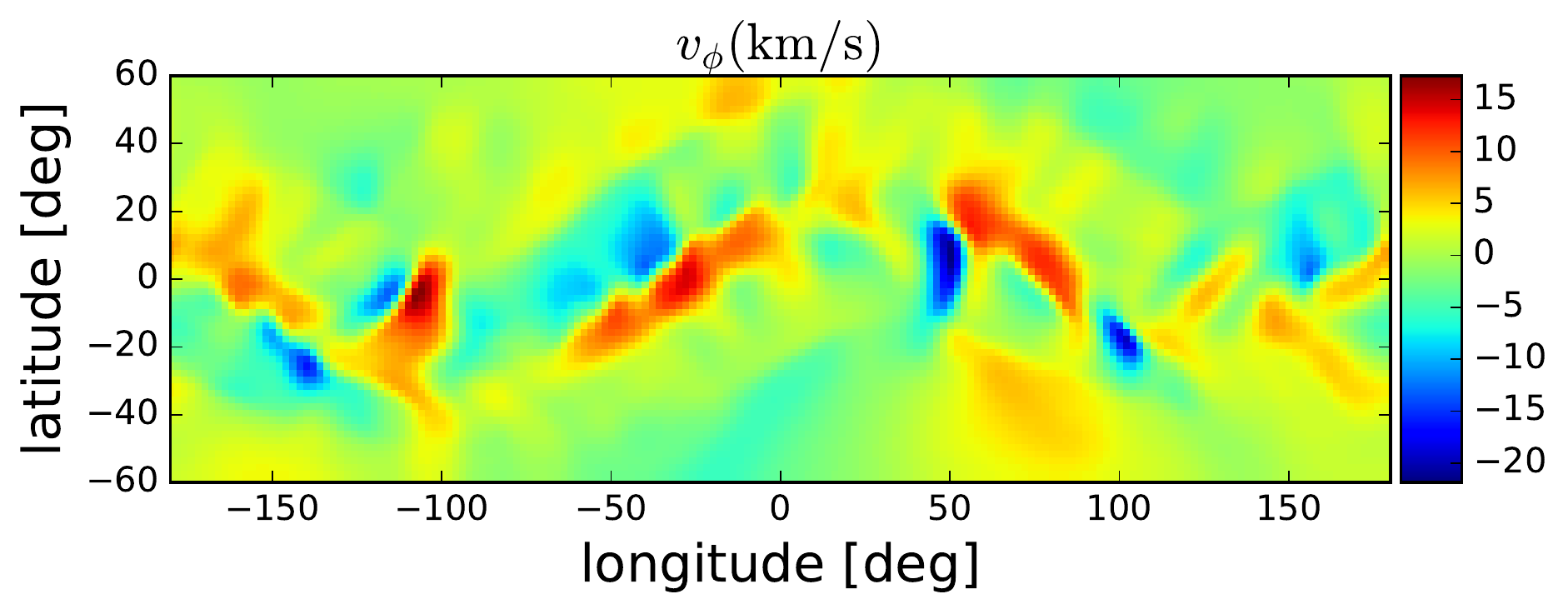}
\includegraphics[trim={0mm 10mm 0mm 0mm},clip=true,width=0.51\textwidth]{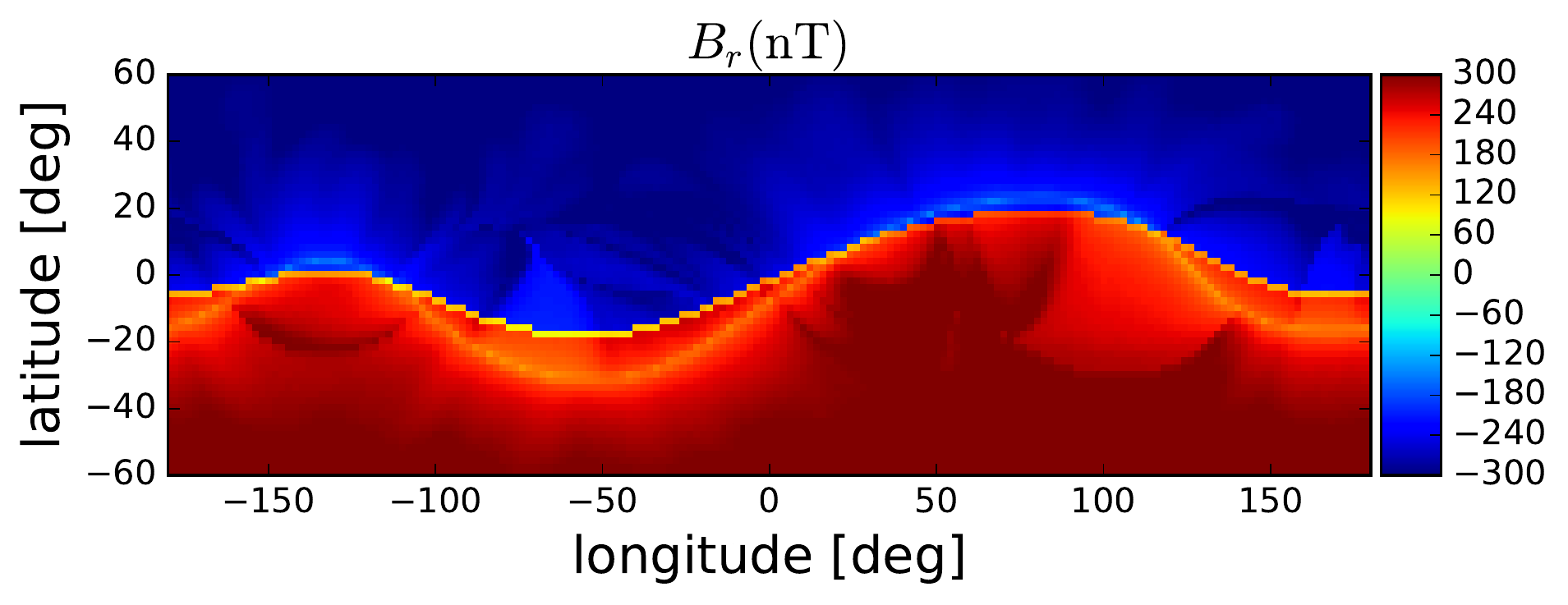}
\includegraphics[trim={10mm 10mm 0mm 0mm},clip=true,width=0.48\textwidth]{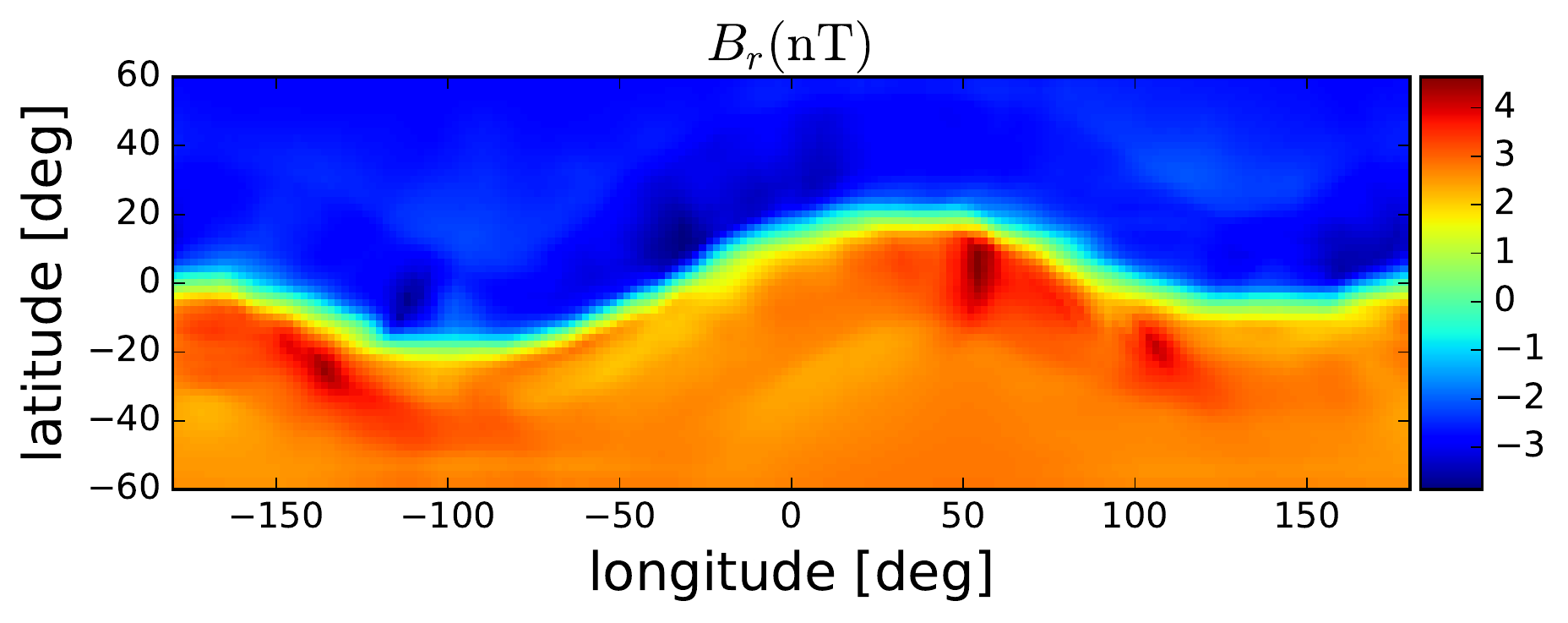}
\includegraphics[trim={0mm 10mm 0mm 0mm},clip=true,width=0.51\textwidth]{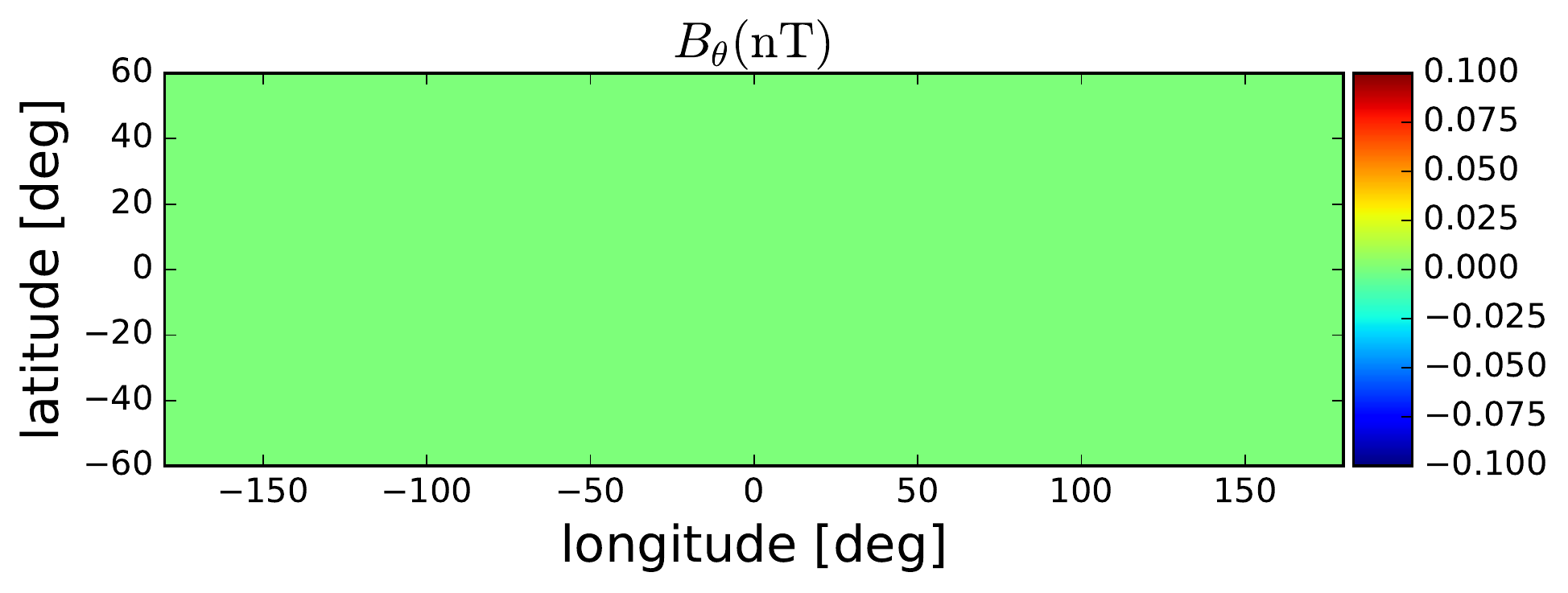}
\includegraphics[trim={10mm 10mm 0mm 0mm},clip=true,width=0.48\textwidth]{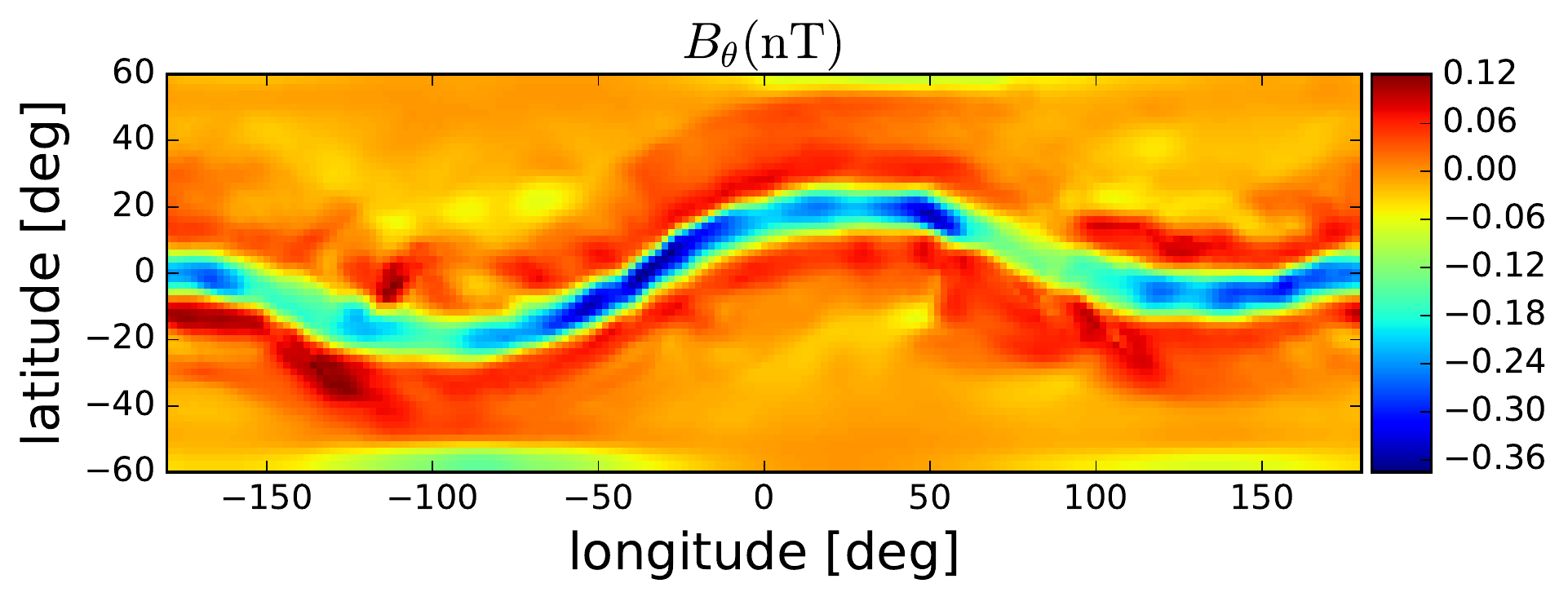}
\includegraphics[trim={0mm 0mm 0mm 0mm},clip=true,width=0.51\textwidth]{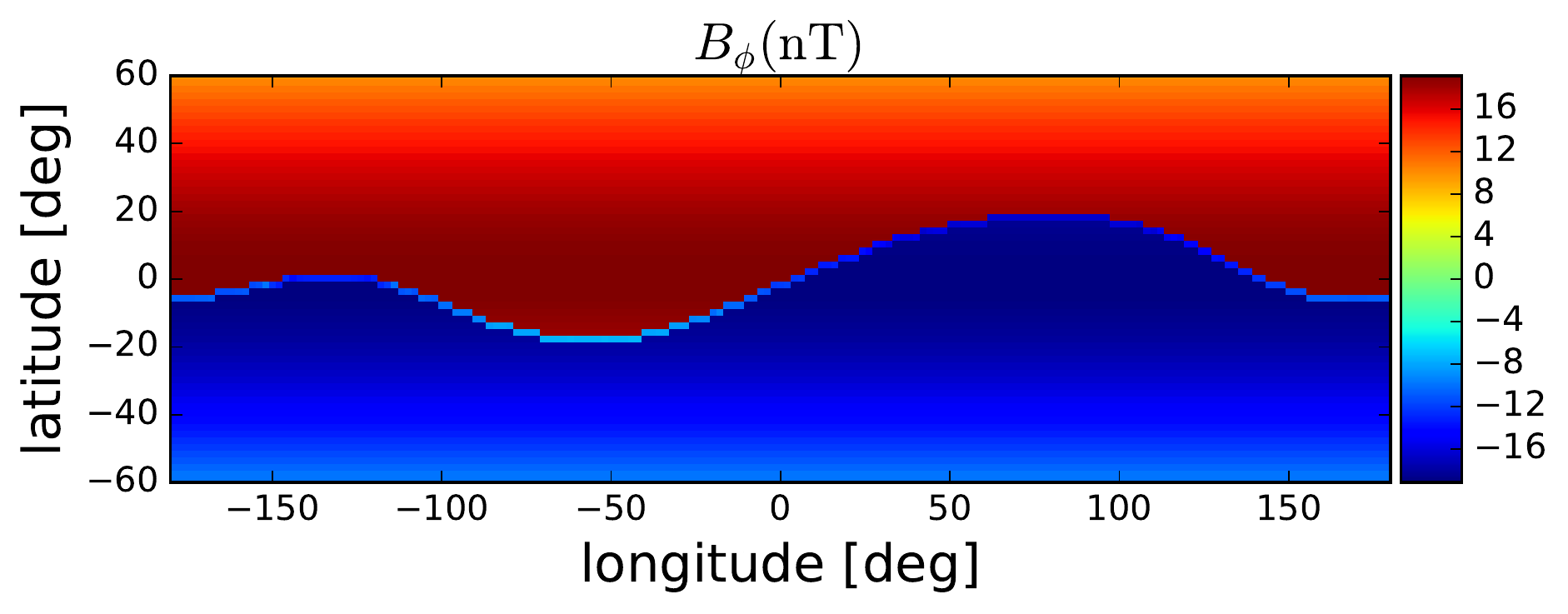}
\includegraphics[trim={10mm 0mm 0mm 0mm},clip=true,width=0.48\textwidth]{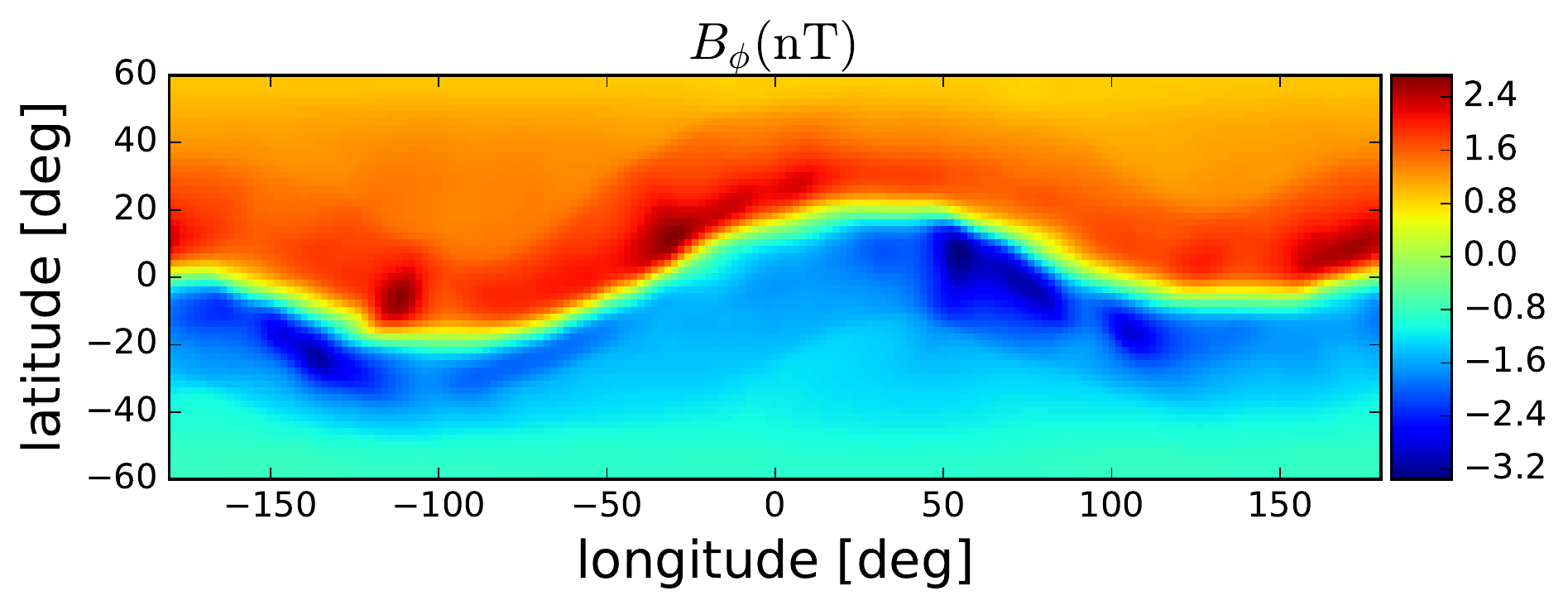}

\caption{EUHFORIA longitudinal and latitudinal variations of the three velocity and the three magnetic field components at 0.1AU (\textit{left panels}) and at 1AU (\textit{right panels}) for a standard magnetogram centered at 10 August 2007.}
\label{fig:eufvrbr}
\end{center}
\end{figure}

In Figure~\ref{fig:eufvrbr}, we show the components of the
velocity and magnetic field in the spherical
($r,\theta,\phi$) basis at two different distances, 0.1AU and 1AU, as
functions of heliospheric longitude and latitude, as calculated by EUHFORIA. We
observe a clear and narrow undulating current sheet showing clear discrimination
between low- and high-latitude solar wind. This undulating sheet is characterized
by lower velocity than other regions at 0.1AU. It separates the outward (southern
hemisphere) and inward (northern hemisphere) magnetic field topologies. As the
solar wind propagates outwards, the interactions between streams of different
speeds become increasingly important with the sharp features at 0.1AU turn
into more diffused, smoothed and extended ones at larger distances. At the
Earth's orbit the current sheet and thus the slow speed region is thicker
covering about $10^\circ$ in latitude. The speed difference at 1AU with respect to
the inner boundary is about $50\mathrm{km\ s^{-1}}$. The $\theta$,
$\phi$-components of the velocity increase in magnitude from zero to about
$\approx 15 \mathrm{km\ s^{-1}}$ at the Earth's orbit, roughly following the
pattern that the slow speeds appear at small latitudes contrary to the high
solar wind speeds. The radial component of the magnetic field decreases by about
2 orders of magnitude from the inner boundary to the orbit of the Earth showing
a broader current sheet ($\approx 5^\circ$), where the magnetic field is close to zero.
The $\theta$-component of the magnetic field increases from zero to about $0.36$
nT in magnitude up to 1AU. On the other hand, the $B_\phi$ decreases by almost a factor of 8
up to 1AU and is at both distances opposite in polarity to the radial magnetic
field, thus having positive polarity at the North Pole and negative at the
South. In all the depicted quantities the rotation of the plasma shows as the
entire structure in the 2D maps moves to the left.

\begin{sidewaysfigure}[htbp]
\begin{center}
\includegraphics[trim={0mm 10mm 0mm 0mm},clip=true,width=0.325\textwidth]{vr_aug_10_2007_01au.pdf}
\includegraphics[trim={10mm 10mm 0mm 0mm},clip=true,width=0.325\textwidth]{vr_aug_10_2007_1au.pdf}
\includegraphics[trim={10mm 10mm 0mm 0mm},clip=true,width=0.325\textwidth]{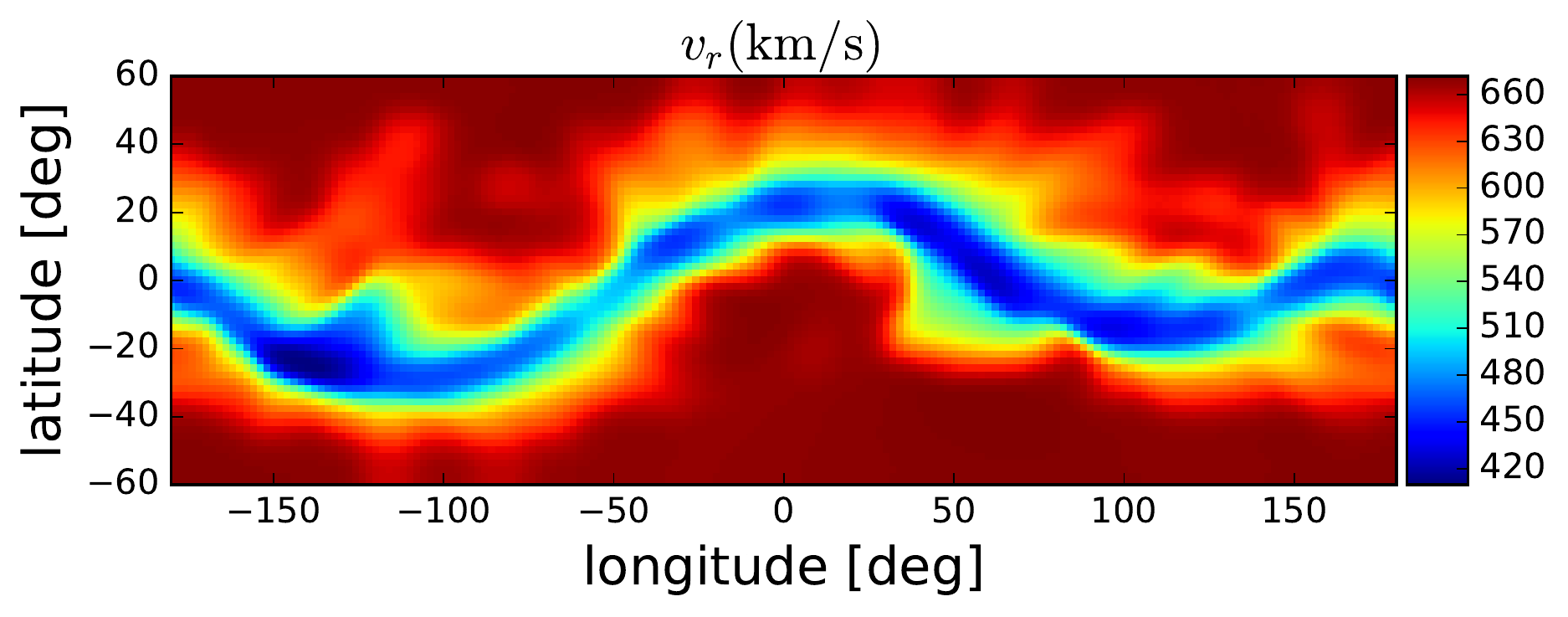}
\includegraphics[trim={0mm 10mm 0mm 0mm},clip=true,width=0.325\textwidth]{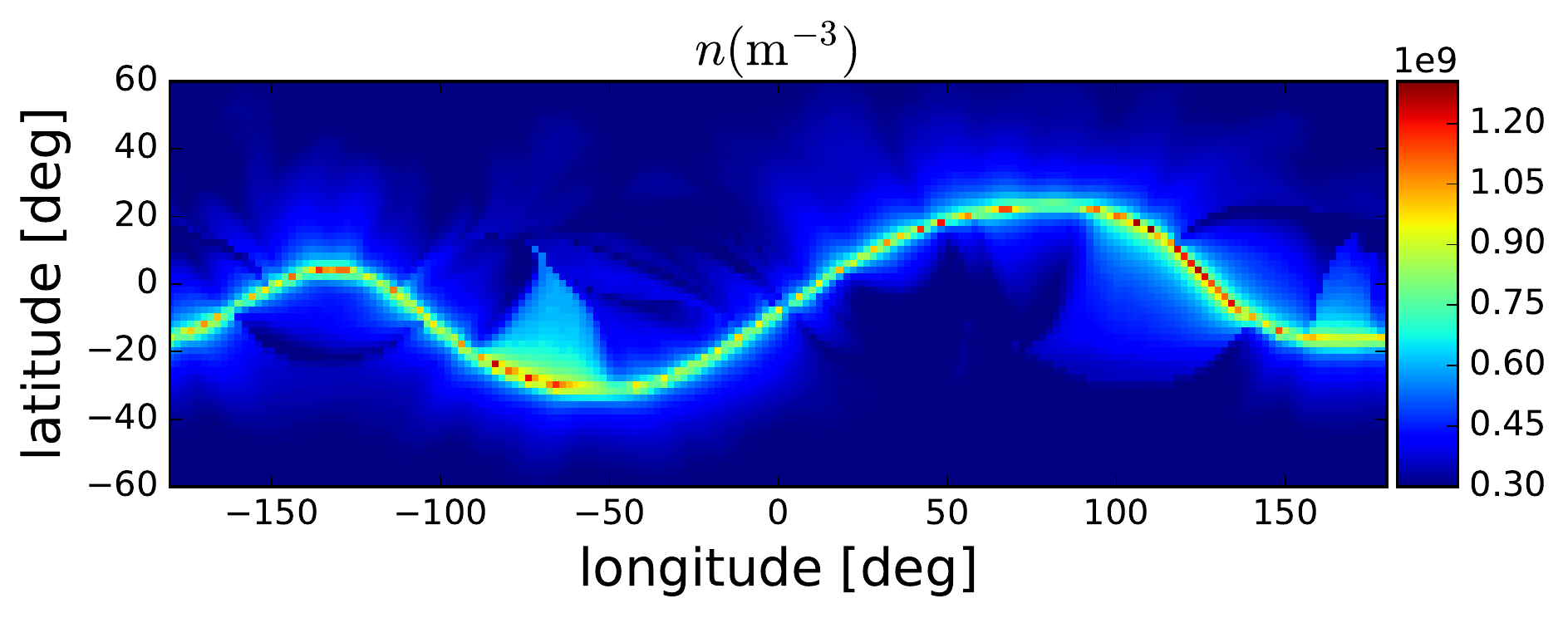}
\includegraphics[trim={10mm 10mm 0mm 0mm},clip=true,width=0.325\textwidth]{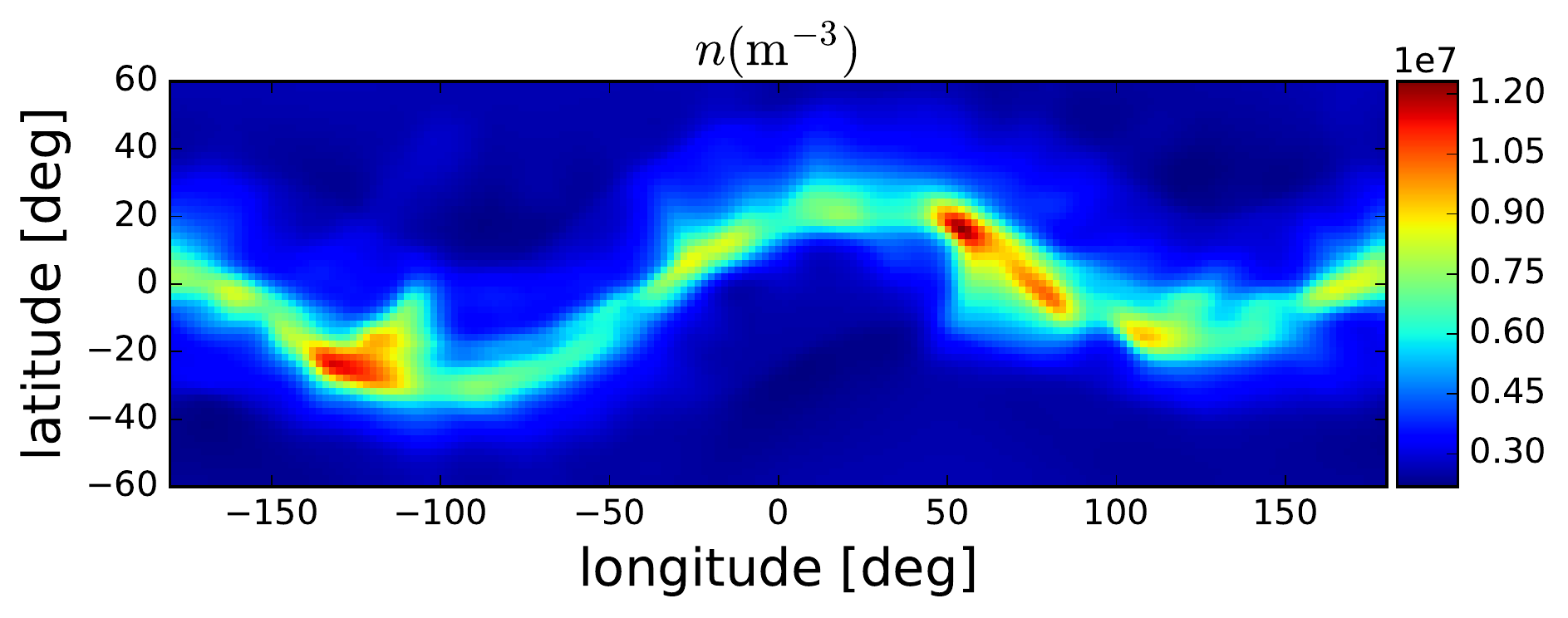}
\includegraphics[trim={10mm 10mm 0mm 0mm},clip=true,width=0.325\textwidth]{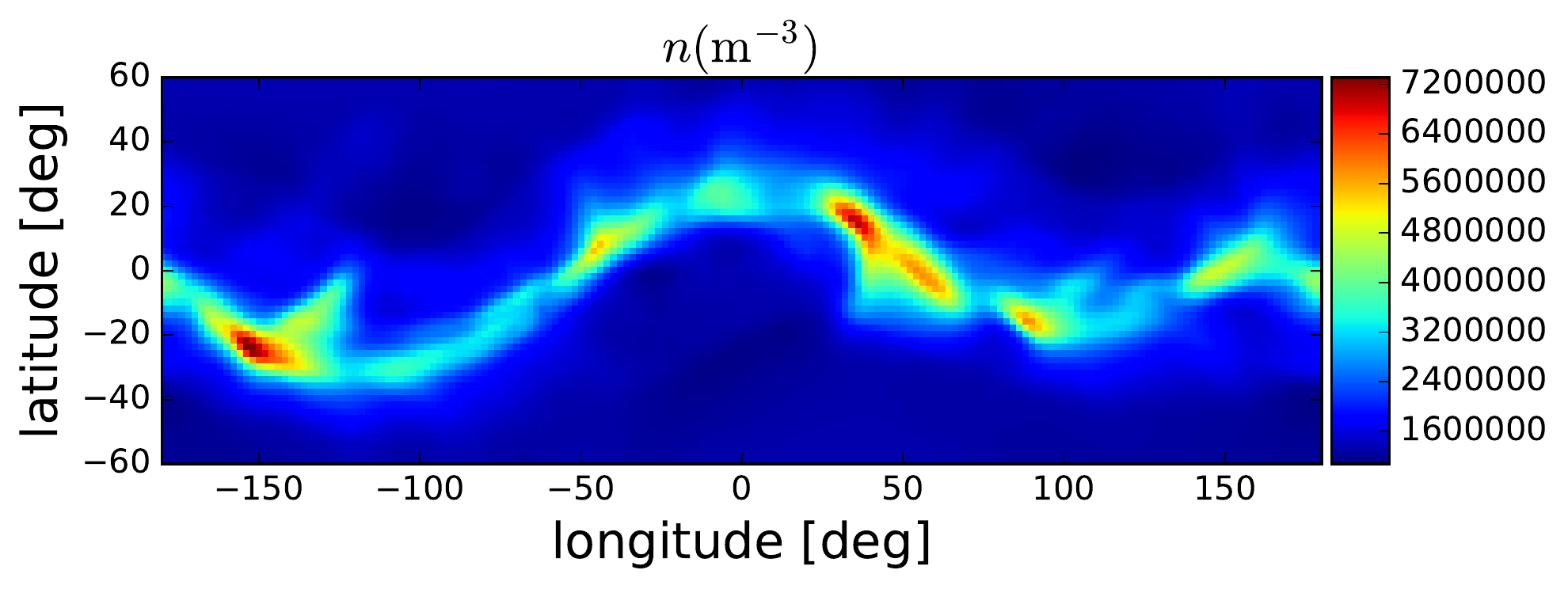}
\includegraphics[trim={0mm 0mm 0mm 0mm},clip=true,width=0.325\textwidth]{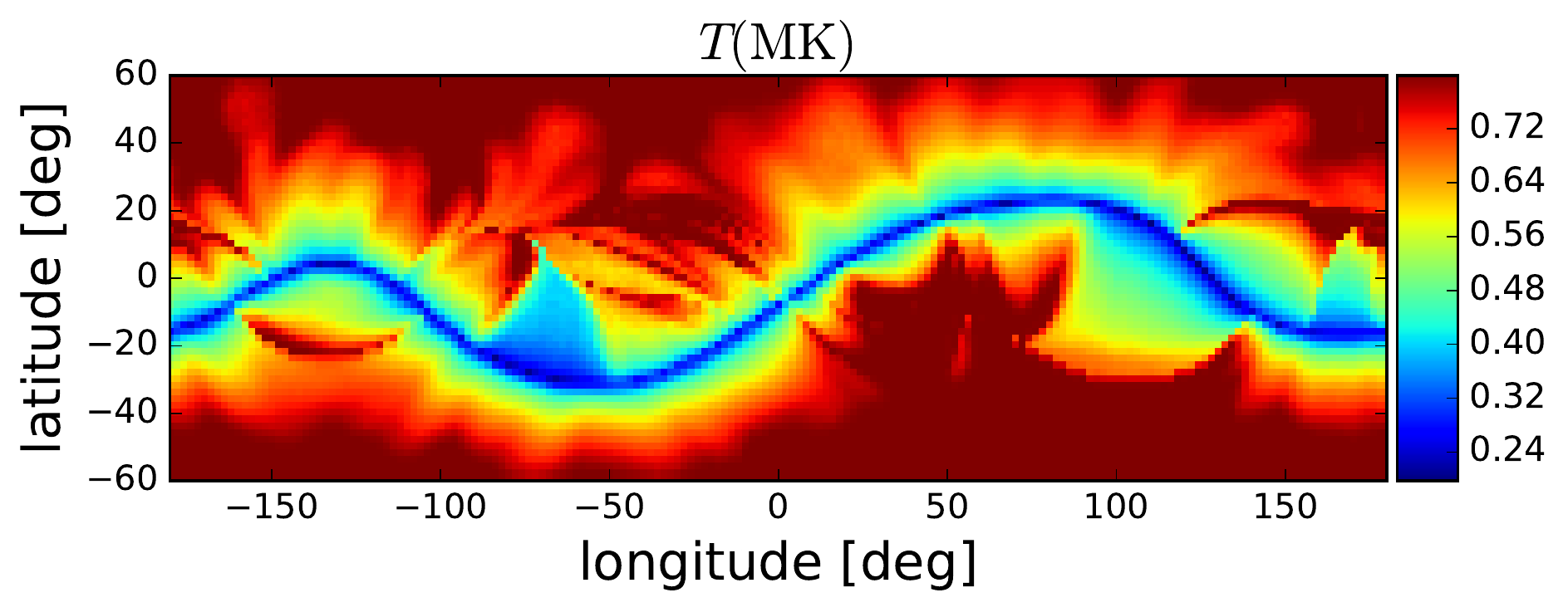}
\includegraphics[trim={10mm 0mm 0mm 0mm},clip=true,width=0.325\textwidth]{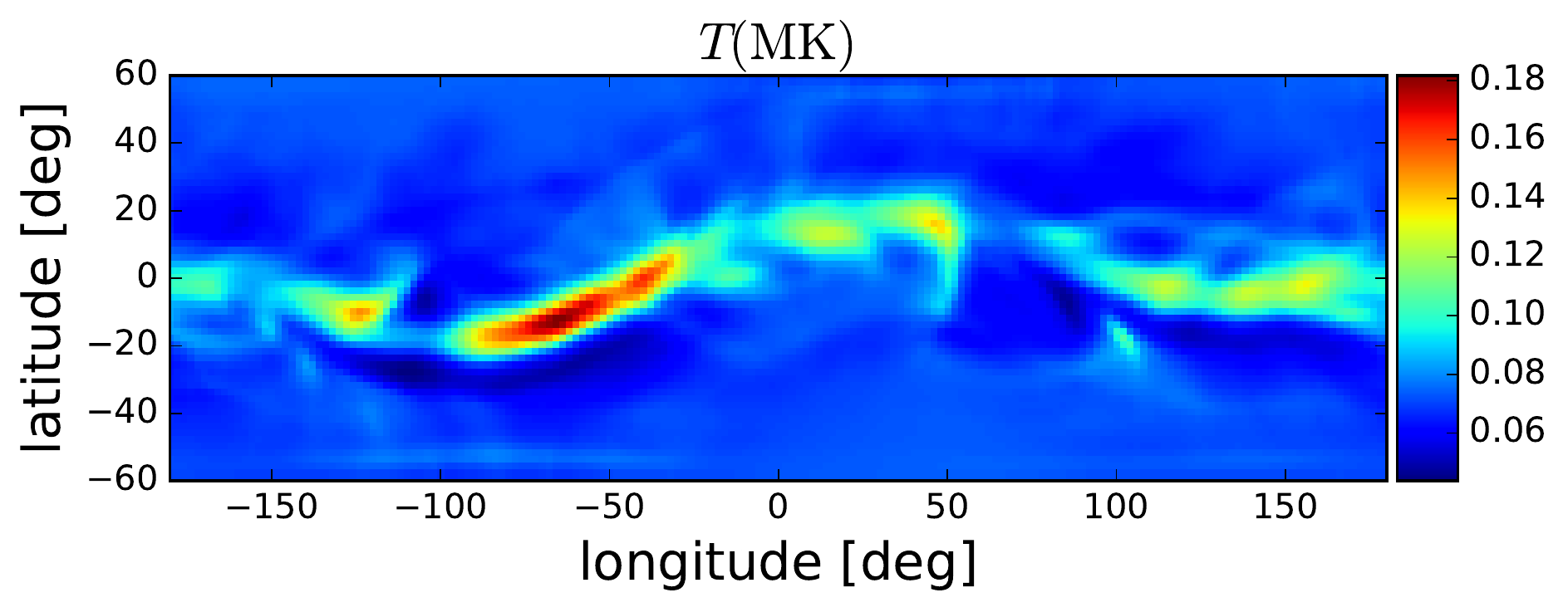}
\includegraphics[trim={10mm 0mm 0mm 0mm},clip=true,width=0.325\textwidth]{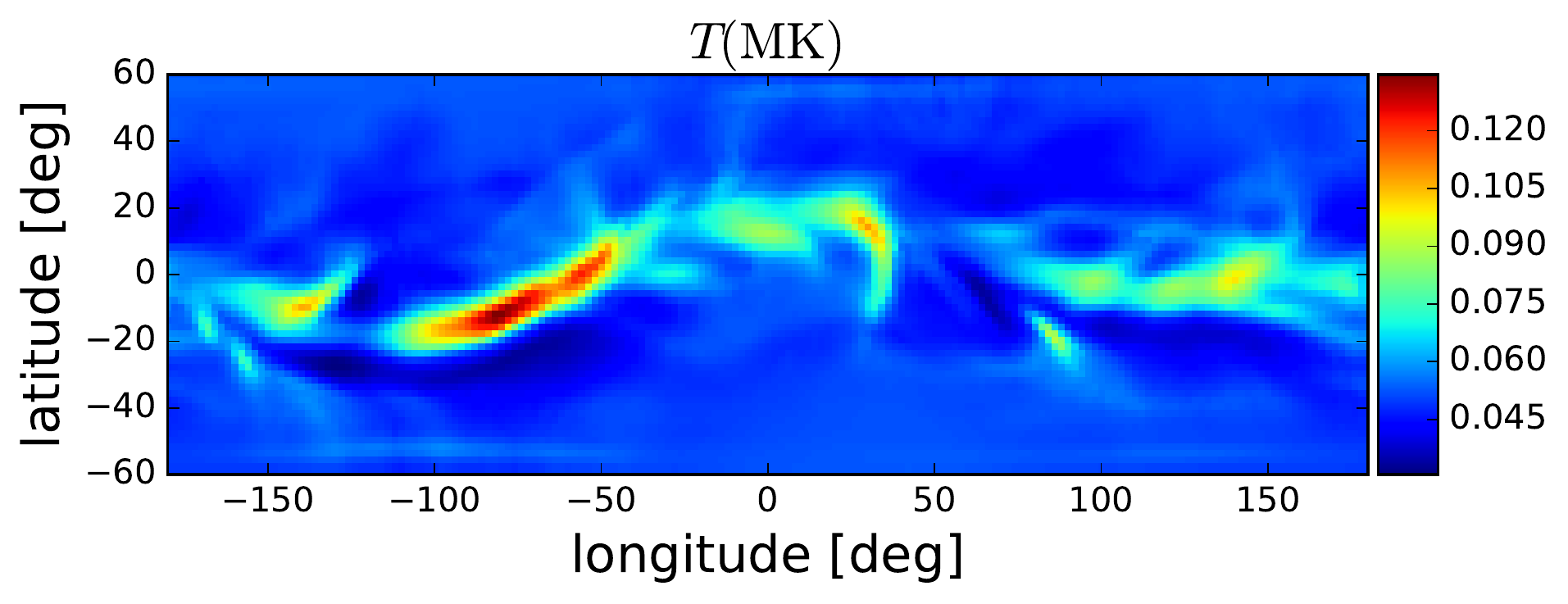}
\caption{EUHFORIA longitudinal and latitudinal variations of the radial velocity (first row), the number density (second row) and the temperature (third row) at 0.1AU (left panels), at 1AU (middle panels) and at 1.4AU (right panels). They can be contrasted with their kinetic counterparts in Figure~\ref{fig:kinslice}.}
\label{fig:eufvrbr3}
\end{center}
\end{sidewaysfigure}
%\clearpage

In Figure~\ref{fig:eufvrbr3}, we show the velocity (first row), the density
(second row) and the temperature (third row) variation at three different
distances, 0.1AU, 1AU, and at the orbit of \textit{Ulysses} at 1.4 AU, \textit{i.e.}\
$\approx 300\mathrm{R}_\odot$, corresponding to left, middle and right columns,
respectively. Most of the acceleration has already taken place at 1AU and only a
small increase is happening above that distance due to the heating implemented
by the reduced polytropic index ($\gamma=1.5$) as the wind flows away from the Sun. The density decreases
by 2 orders of magnitude from the inner boundary of the MHD simulation to the
Earth's orbit and only by 50\% from there onwards up to 1.4AU. The current
sheet, seen as the high density structure that appears in the middle of the
Figures in the second row seems to have expanded from $2^\circ$ at the inner
boundary to about $10^\circ$ from there on, while it diffuses. The temperature
decreases by a factor 8 from the inner boundary up to 1AU and only by 30\% up to
\textit{Ulysses}' orbit. There is a temperature reversal captured, in the sense that
while at 0.1AU the equatorial region appears cold and the poles hot, the
opposite is happening from 1AU onwards, with the temperature shows a peak at a
longitude of $-100^\circ$. The current sheet region gets very diffused outwards and
from the Earth's orbit outwards appears discontinuous in this temperature view
of the expanding plasma.

\subsubsection{Interfacing}
%%%%%%%%%%%%%%%%%%%% Interfacing %%%%%%%%%%%%%%%%%%%%%%%%%%%%
In order to interface the two models at 0.1AU, we run the kinetic model up to
$21.5\mathrm{R}_\odot$ with $r_0=r_s=2.5\mathrm{R}_\odot$, $T_e=1\mathrm{MK}$, $T_p=1\mathrm{MK}$,
using 600 $\kappa$-indexes in the range [2,8] with step 0.01 and with
$n_e=n_p=3\times10^{10}\mathrm{m^{-3}}$~\citep{Lamy03}.
With the results we create a matrix with
solar wind speeds at $21.5\mathrm{R}_\odot$ and by comparison with the EUHFORIA
results for $v_r$, we estimate the appropriate $\kappa$ for every speed at the
internal boundary of the MHD run. We obtain a 2D map of $N_\theta\times
N_\phi$ values of $\kappa(\theta,\phi)$ each corresponding to a field line. 
From the matrix, we also compare the number density given by EUHFORIA at the same distance (0.1AU) with the number density given by the kinetic model and get a scaling factor that we assume to be valid throughout all the considered radial distances \citep[see Appendix A of][]{Lamy03}.
Thereby, we can estimate the appropriate initial density at the exobase that would give us the same density with EUHFORIA at 0.1AU. 
For the temperatures, the relation is more complicated, but \cite{Lamy03} have demonstrated that the temperature does not affect the kinetic moments as much as the $\kappa$-indexes and the exobase altitude even for extreme changes (1\,--\,2MK) and thus for convenience we take $T_e=T_p=1$MK for all the latitudes and longitudes at $r_0=2.5\mathrm{R}_\odot$ and focus instead on the $\kappa$ and the density parameters. Note that EUHFORIA is not solving the MHD equations below 0.1AU,
while the kinetic model provides results at any distance above the exobase and
especially in the crucial region close to the Sun where the solar wind is being
accelerated.
%%%%%%%%%%%%%%%%%%%%%%%%%%%%%%%%%%%%%%%%%%%%%%%%%%%%%%%

\subsubsection{Kinetic Model}

\begin{sidewaysfigure}[htbp]
\begin{center}
\includegraphics[trim={0mm 10mm 0mm 0mm},clip=true,width=0.325\textwidth]{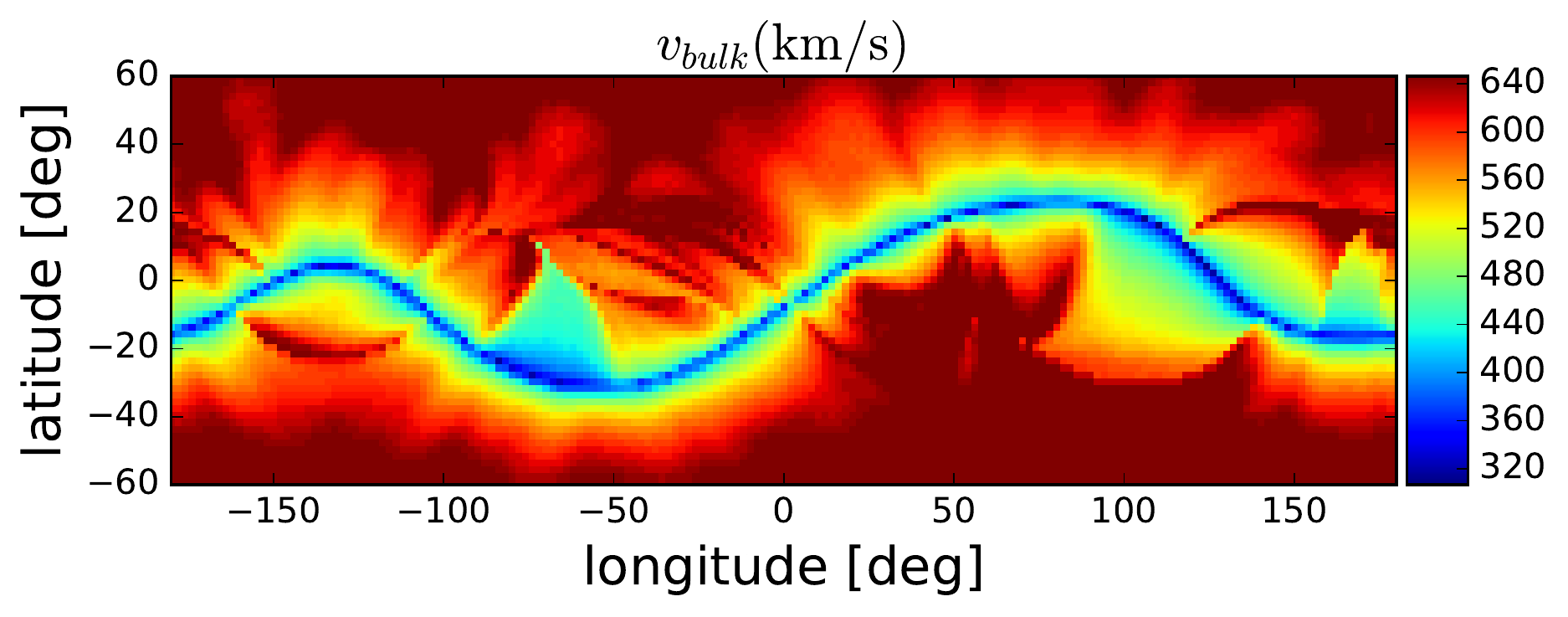}
\includegraphics[trim={10mm 10mm 0mm 0mm},clip=true,width=0.325\textwidth]{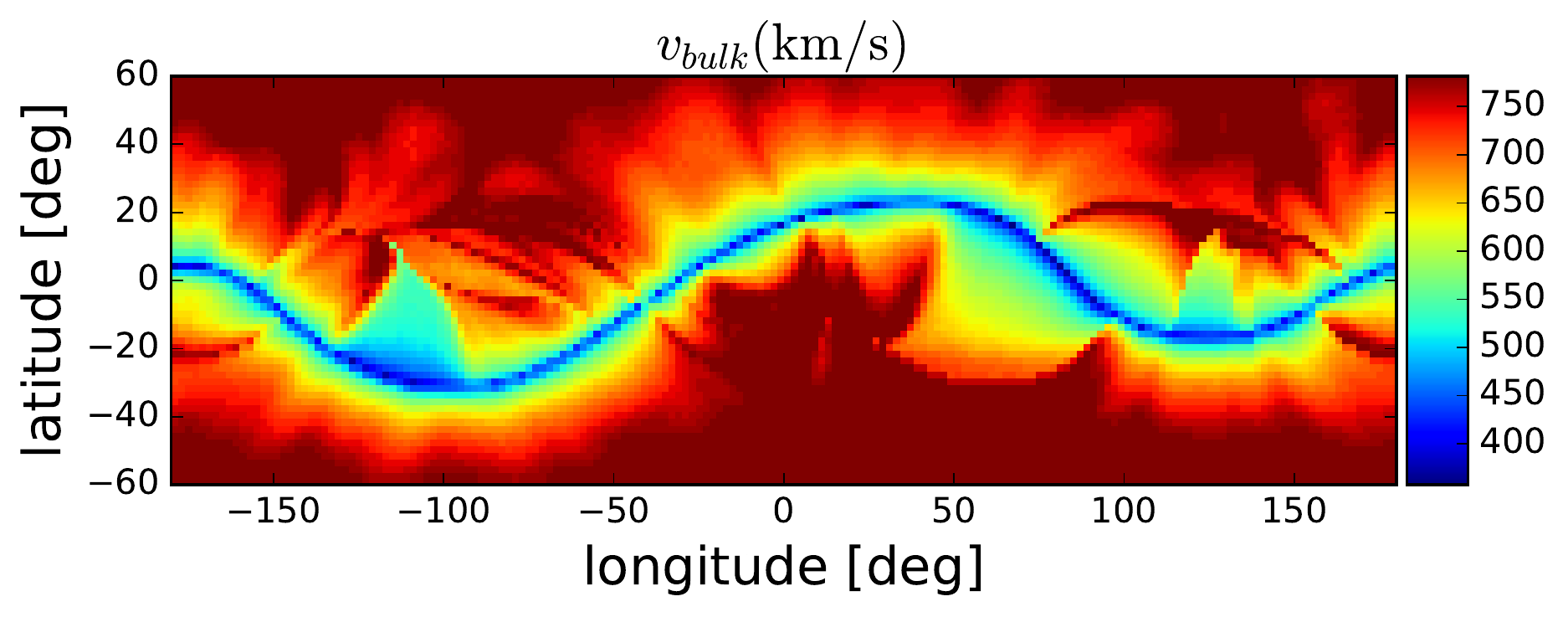}
\includegraphics[trim={10mm 10mm 0mm 0mm},clip=true,width=0.325\textwidth]{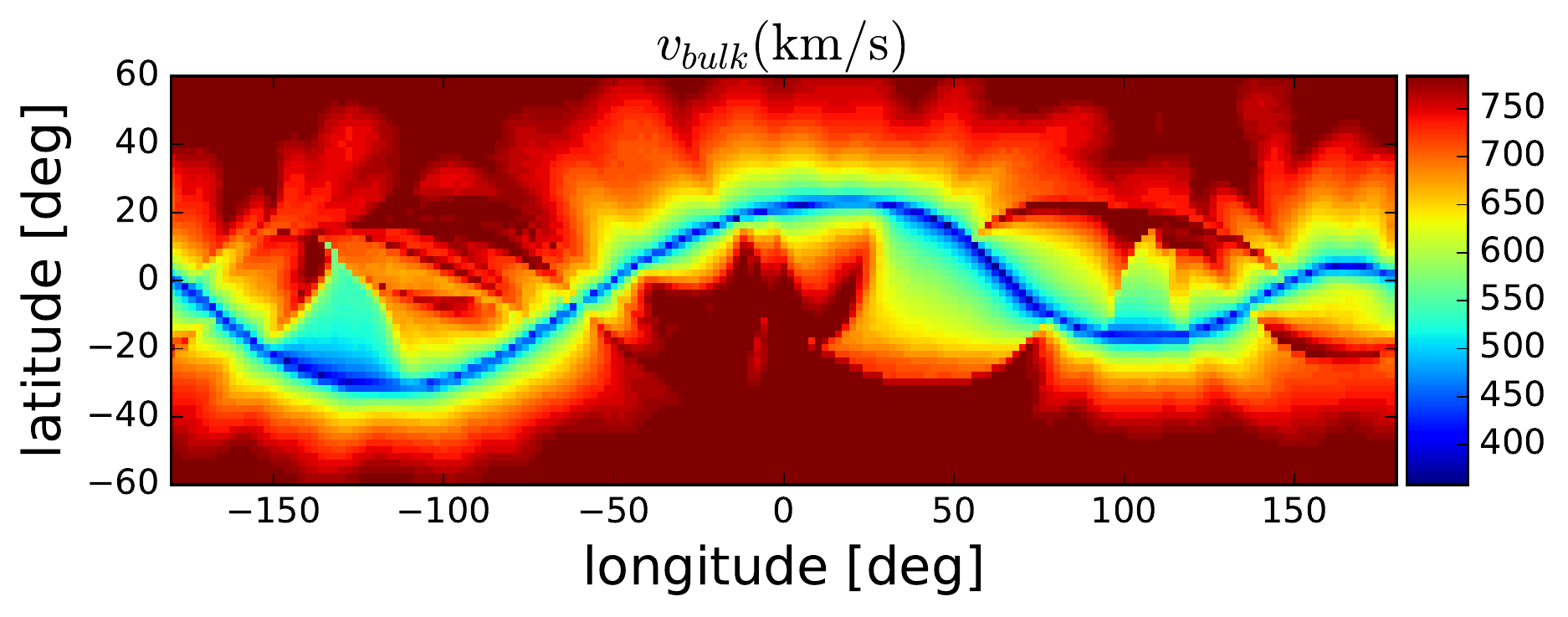}
\includegraphics[trim={0mm 10mm 0mm 0mm},clip=true,width=0.325\textwidth]{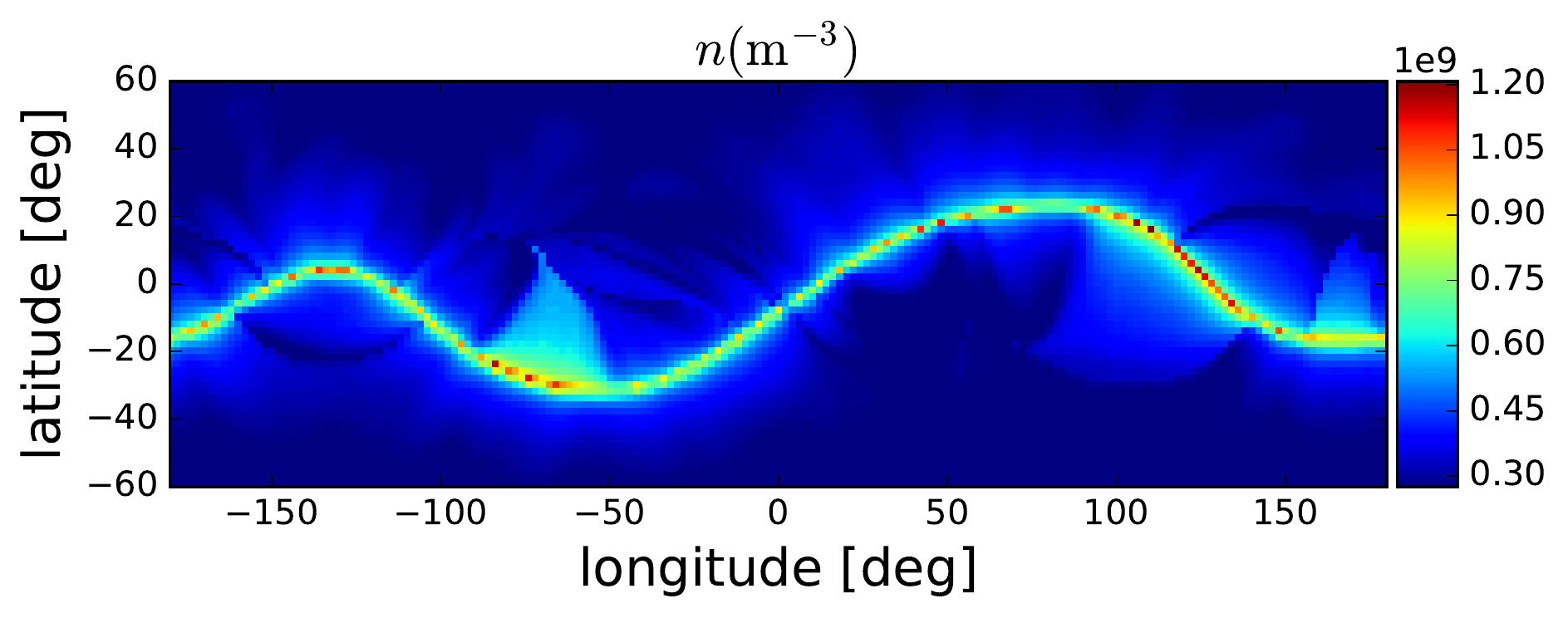}
\includegraphics[trim={10mm 10mm 0mm 0mm},clip=true,width=0.325\textwidth]{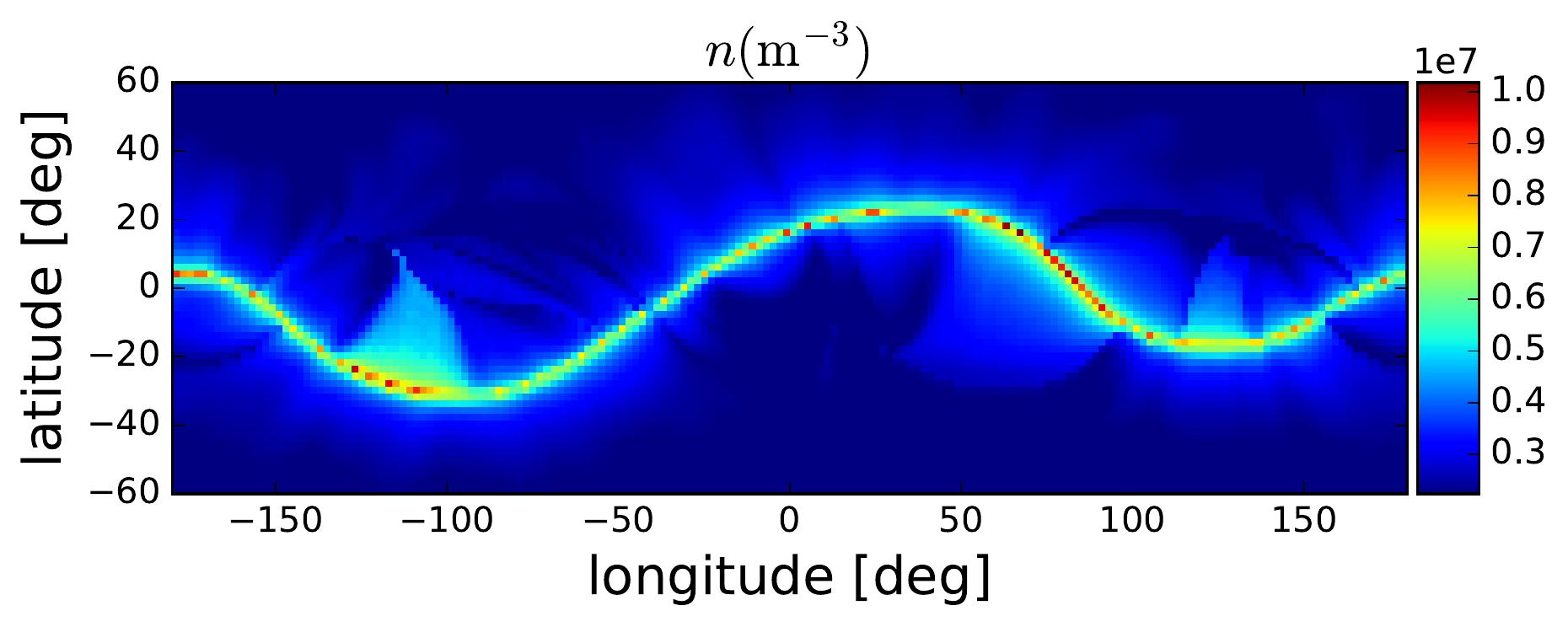}
\includegraphics[trim={10mm 10mm 0mm 0mm},clip=true,width=0.325\textwidth]{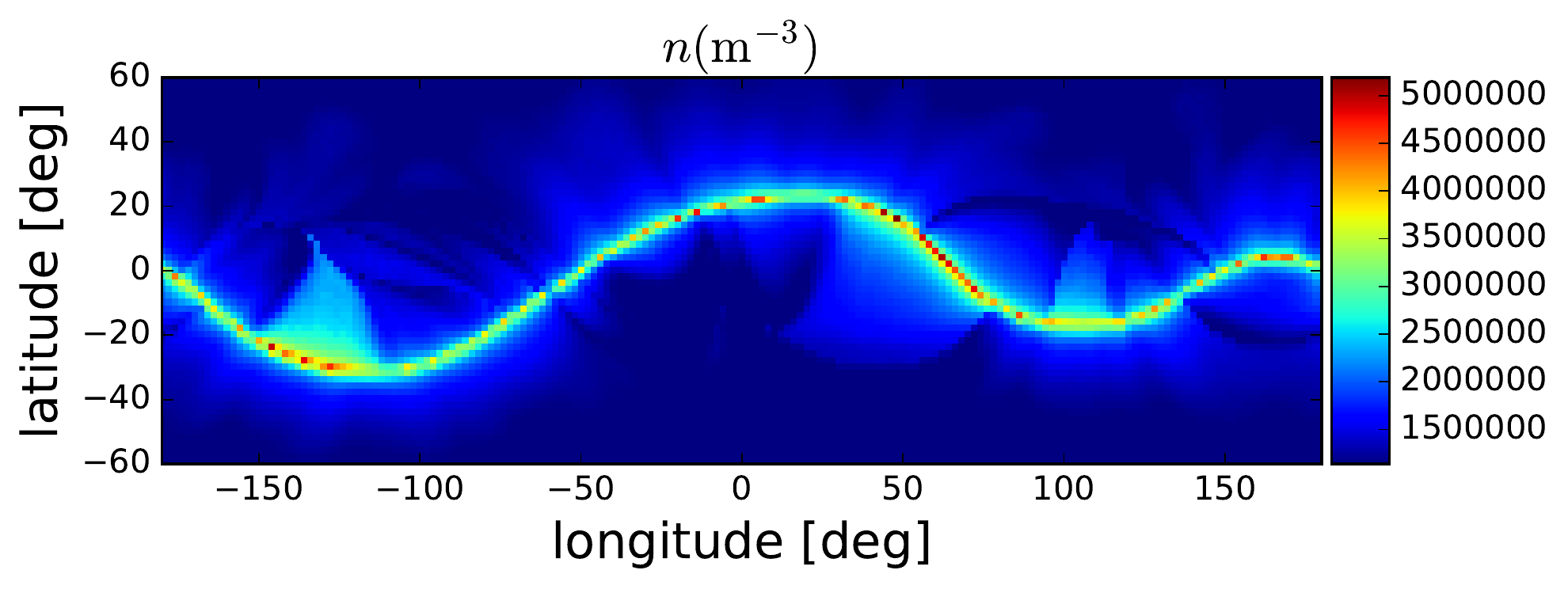}
\includegraphics[trim={0mm 10mm 0mm 0mm},clip=true,width=0.325\textwidth]{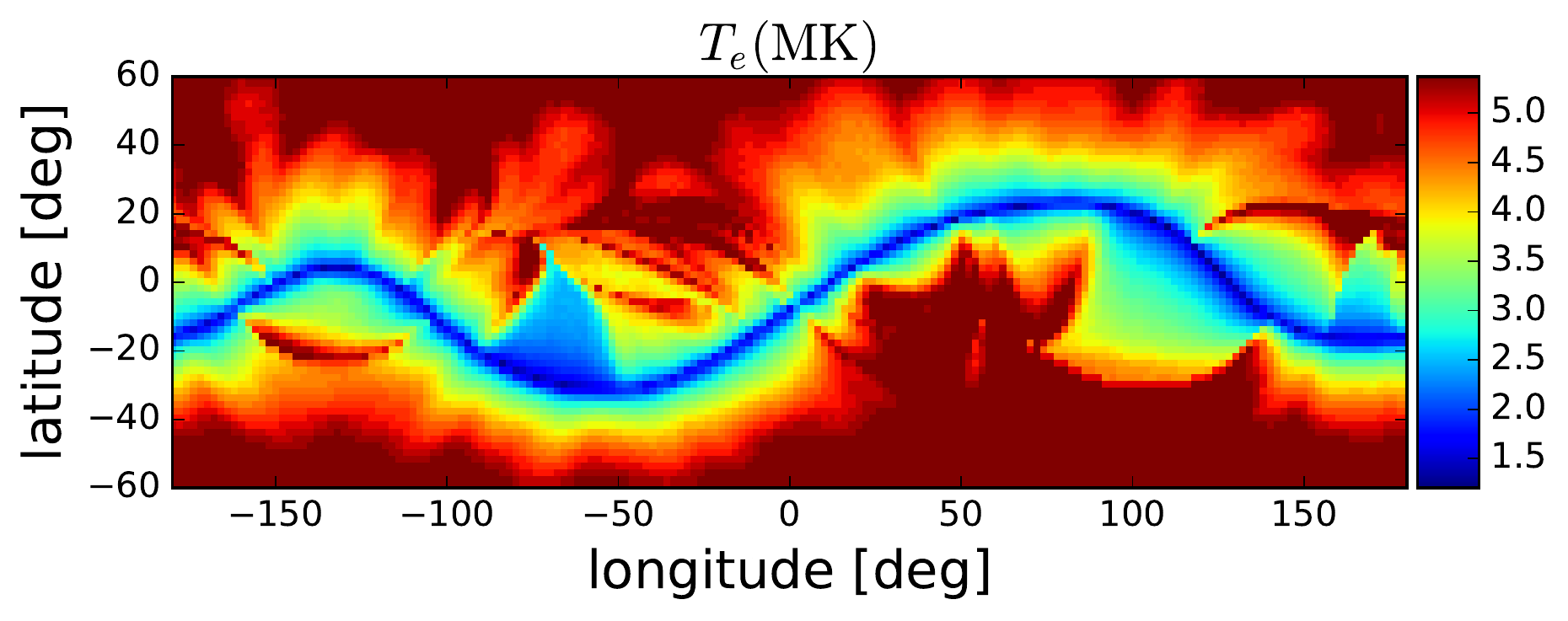}
\includegraphics[trim={10mm 10mm 0mm 0mm},clip=true,width=0.325\textwidth]{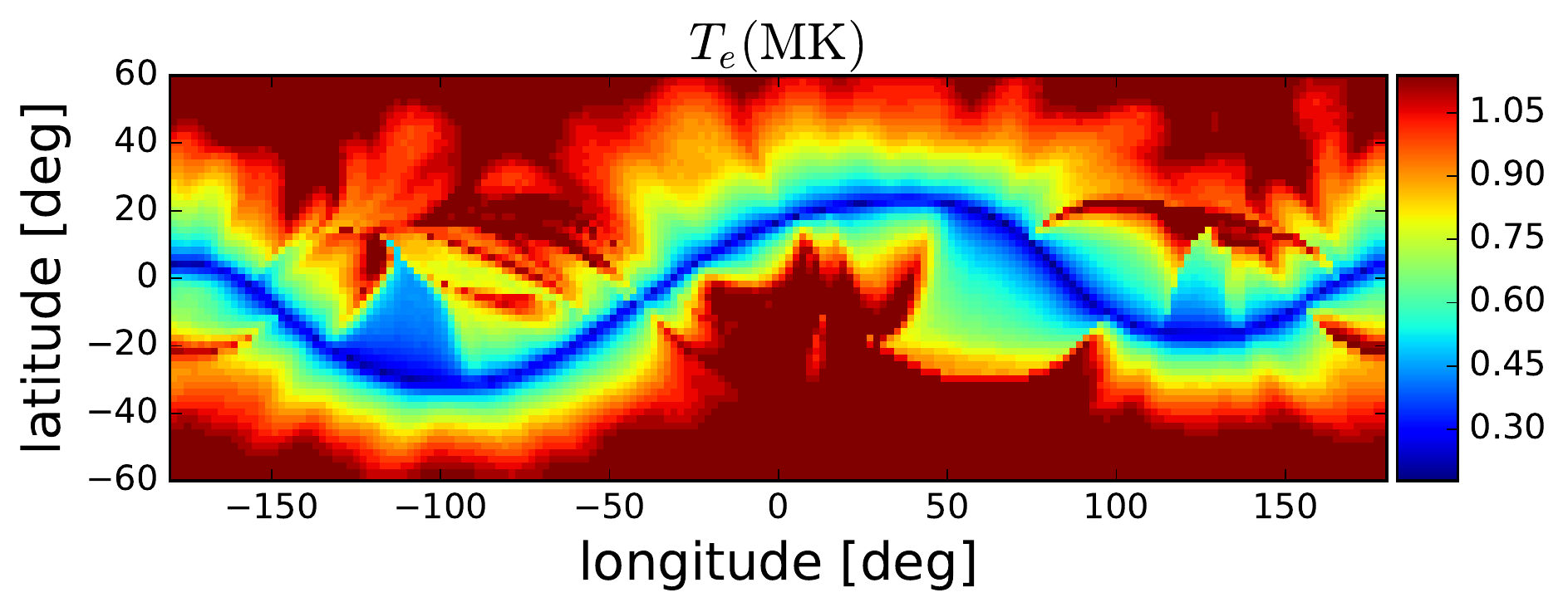}
\includegraphics[trim={10mm 10mm 0mm 0mm},clip=true,width=0.325\textwidth]{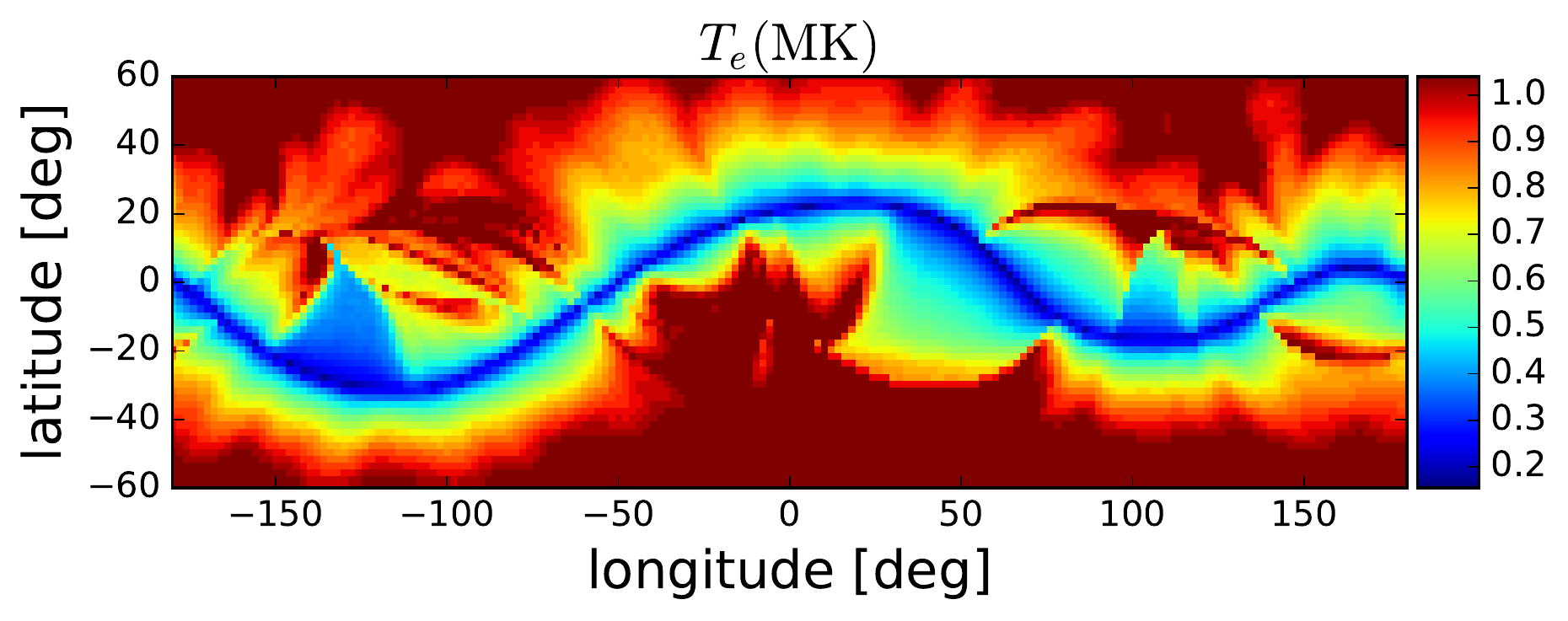}
\includegraphics[trim={0mm 0mm 0mm 0mm},clip=true,width=0.325\textwidth]{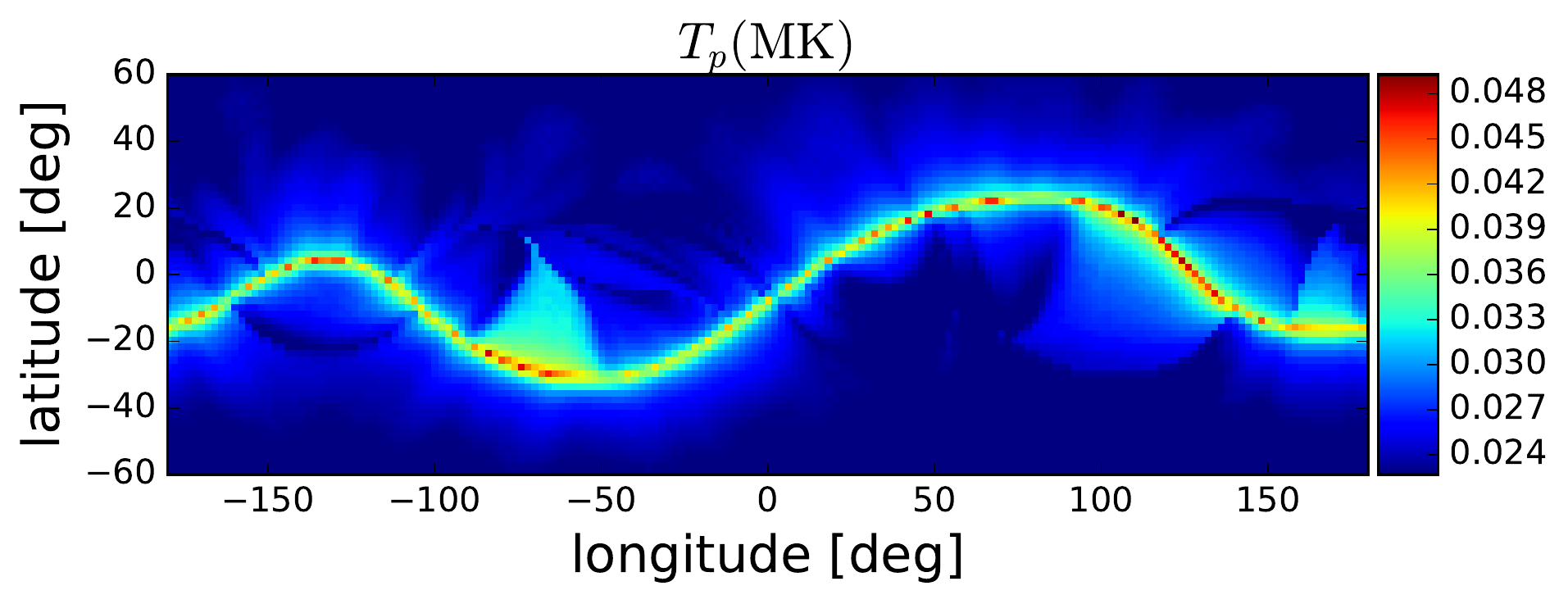}
\includegraphics[trim={10mm 0mm 0mm 0mm},clip=true,width=0.325\textwidth]{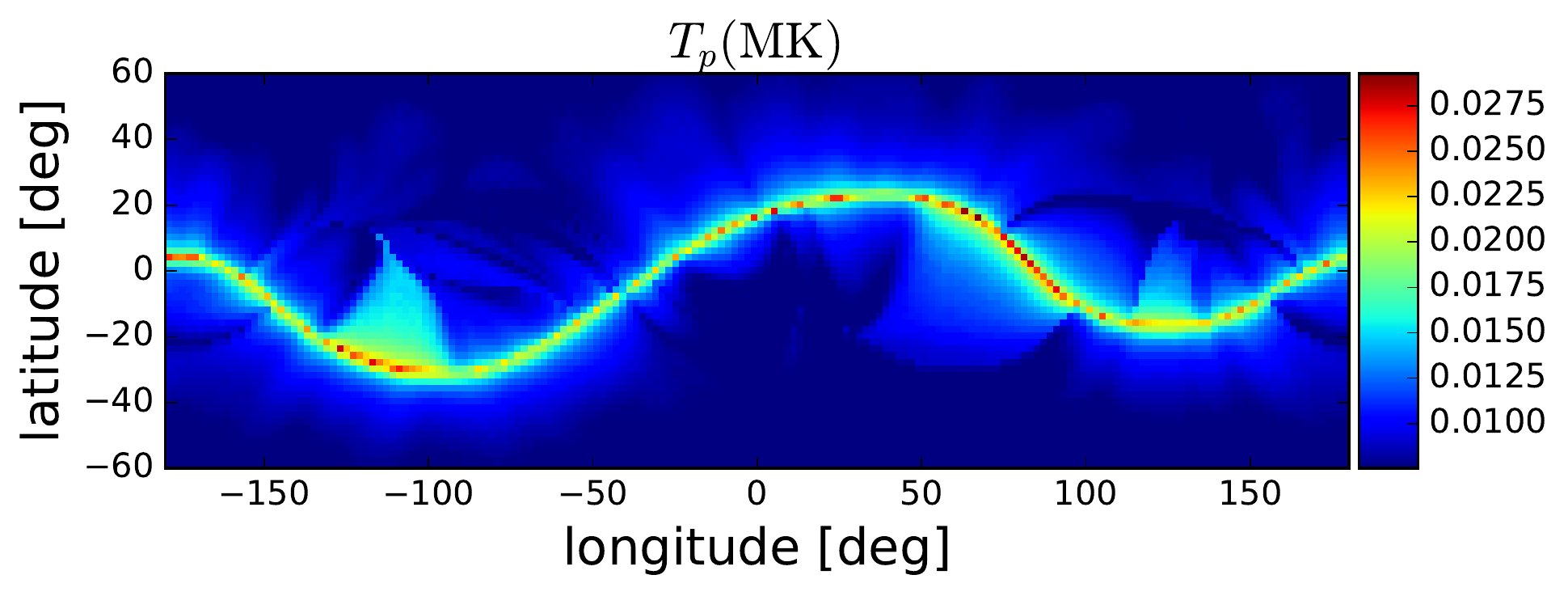}
\includegraphics[trim={10mm 0mm 0mm 0mm},clip=true,width=0.325\textwidth]{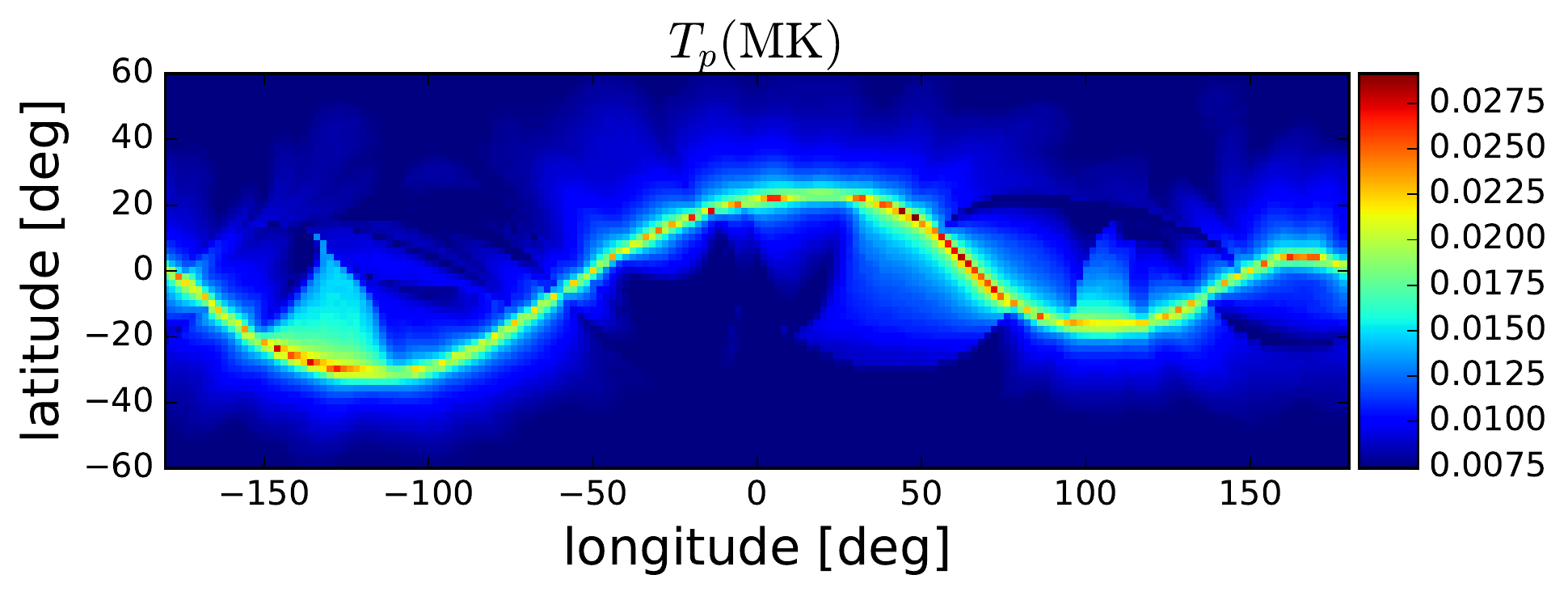}
\caption{Kinetic results in the form of color maps in heliographic longitude and latitude from top to bottom rows: of the bulk speed, the number density, the electron and proton temperatures of the solar wind respectively at 0.1AU (\textit{left panels}), at 1AU (\textit{middle panels}) and at \textit{Ulysses}' orbit 1.4AU (\textit{right panels}). These can be contrasted directly with the MHD counterparts in Figure~\ref{fig:eufvrbr3}.}
\label{fig:kinslice}
\end{center}
\end{sidewaysfigure}

Similar to Figure~\ref{fig:eufvrbr3}, in Figure~\ref{fig:kinslice}, we show the velocity (first row), the density
(second row), the electron temperature (third row) and the proton temperature
(fourth row) at three different distances, 0.1AU, 1AU, and at the
orbit of \textit{Ulysses} at 1.4AU, \textit{i.e.}\ $\approx 300\mathrm{R}_\odot$, corresponding to the
left, middle and right columns, respectively. Unlike the MHD results discussed
previously, in the kinetic approach the sharp structures that appear at the
interfacing boundary (0.1AU) remain unchanged as the plasma moves
outwards, as we don't account for stream interactions. In other words,
neighboring streams with different speeds will not interact as they
propagate and they will keep the same topological features up to large radial
distances. The $\kappa$ indexes corresponding to the kinetic velocities of this
simulation lie in the range $(2,4)$. The bulk speed accelerates more than in the
MHD case, reaching terminal speeds about 100 $\mathrm{km\ s^{-1}}$ higher. Thus
in the kinetic approach, where the acceleration is self-consistent and due to
the induced electric field, the acceleration is more efficient than in the MHD
approach, where semi-empirical schemes are used to accelerate the wind.
Similarly to the MHD case, the terminal speed is reached at 1AU and from there
on the acceleration is very slow up to 1.4AU. The density decreases faster by
20\% at 1AU, to reach the orbit of \textit{Ulysses} 30\% smaller than the MHD case. The
temperatures of the electron and proton populations are depicted in the last two
rows. We have started at the exobase by setting both temperatures equal to 1MK
independent of longitude and latitude. The electron temperature drops by a
factor 5 at the orbit of the Earth and it doesn't change much up to 1.4AU, while
the proton temperature decreases by a factor 2 and remains about the same up to
the orbit of \textit{Ulysses}. The temperature of the MHD plasma is always in between the
electron and proton temperatures, being one order of magnitude smaller than the
electron and one order of magnitude larger than the proton temperature. The MHD
temperature is not the average between the two particle species temperatures,
since the two species are not in thermodynamic equilibrium.

%\clearpage

%%%%%%%%%%%%%%%%%%%%%%%%%%%%%%%%%%%%%%%%%%%%%%%%%%%%%%%%%%%%%%
\subsubsection{Models \textit{Versus} Observations}

%%%%%%%%%%%%%%%% OMNI 2007 %%%%%%%%%%%%%%%%
\begin{figure}[htbp]
\begin{center}
\includegraphics[width=0.49\textwidth]{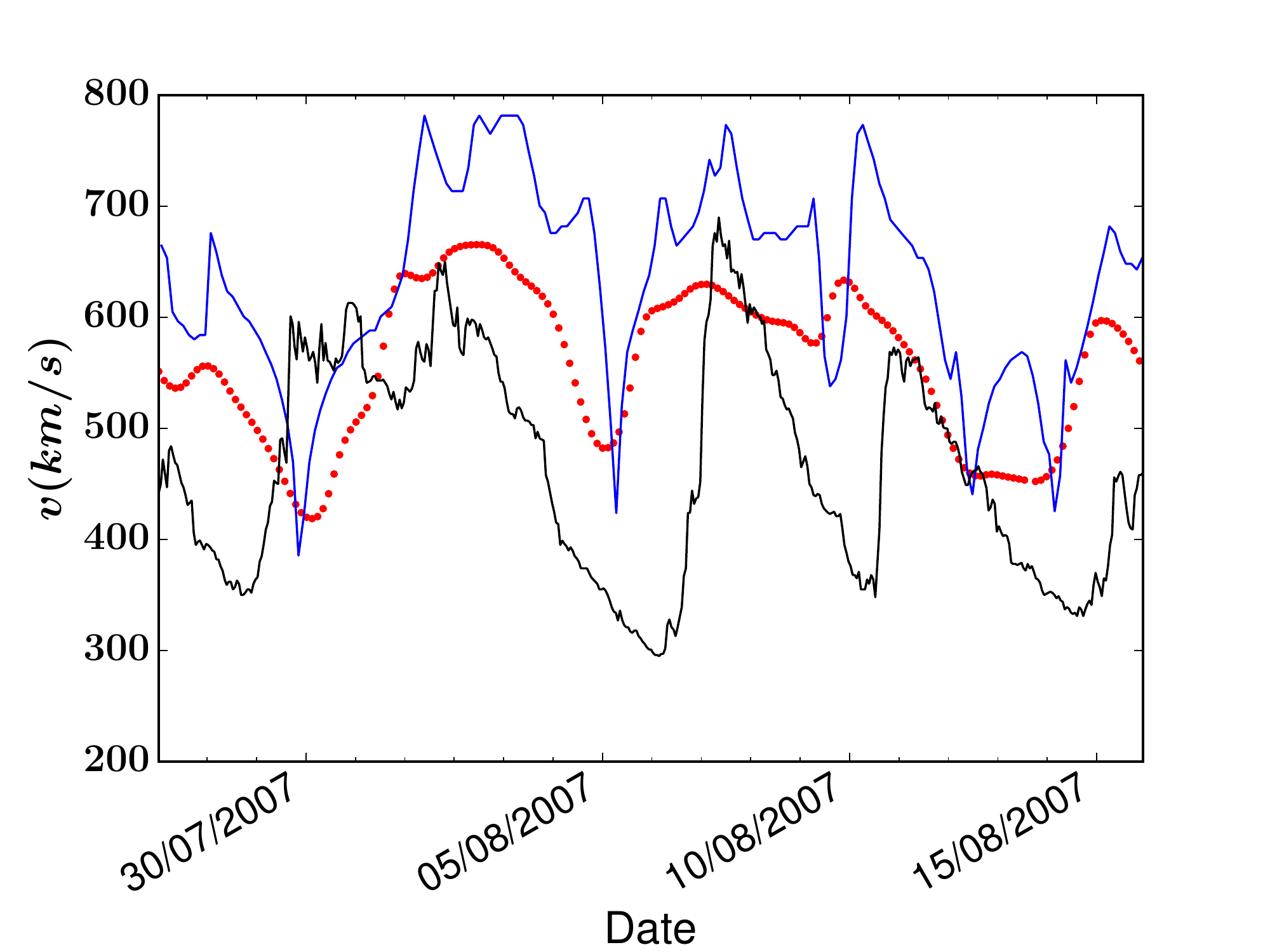}\\
\includegraphics[width=0.49\textwidth]{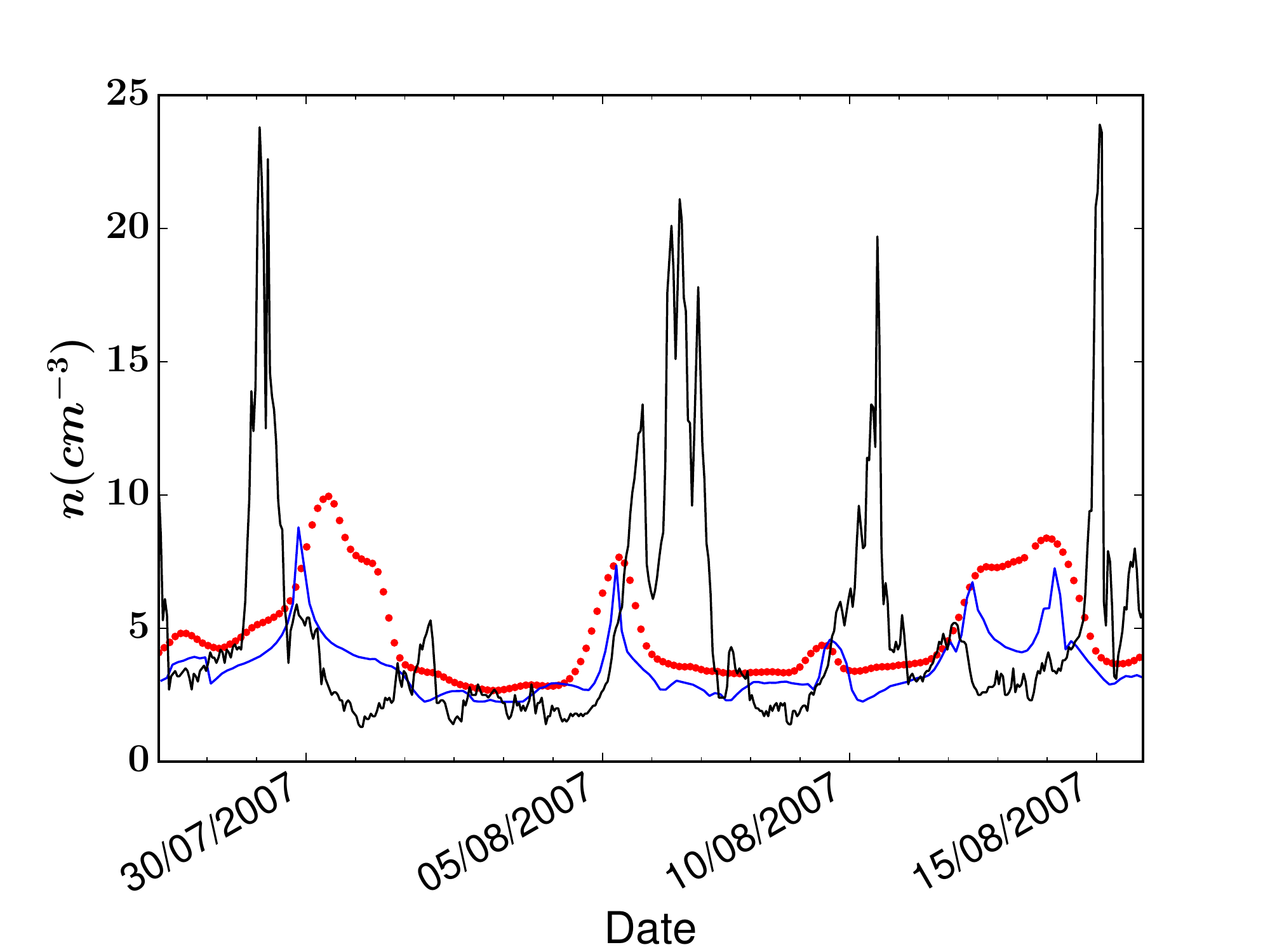}
\includegraphics[width=0.49\textwidth]{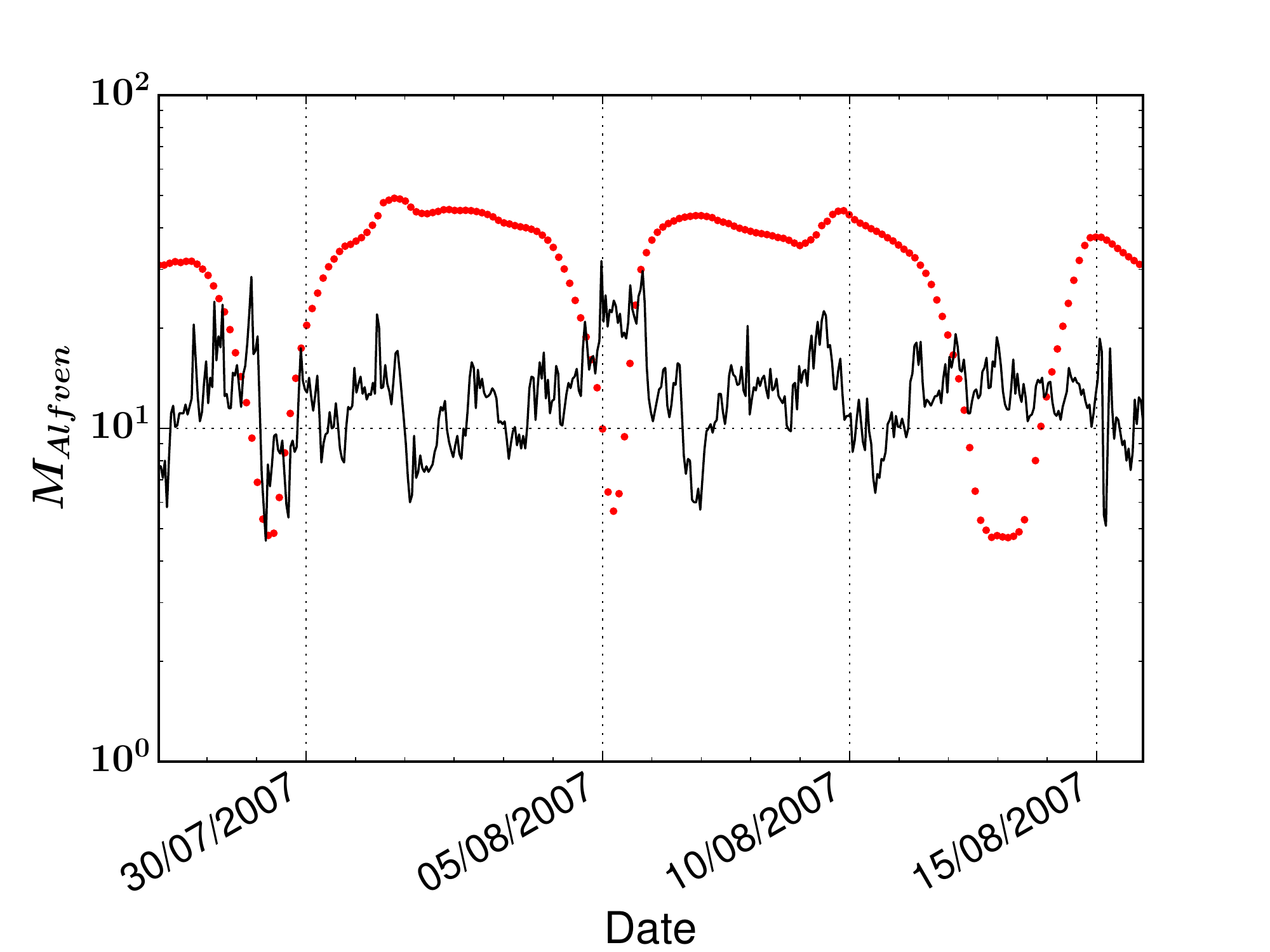}
\includegraphics[width=0.49\textwidth]{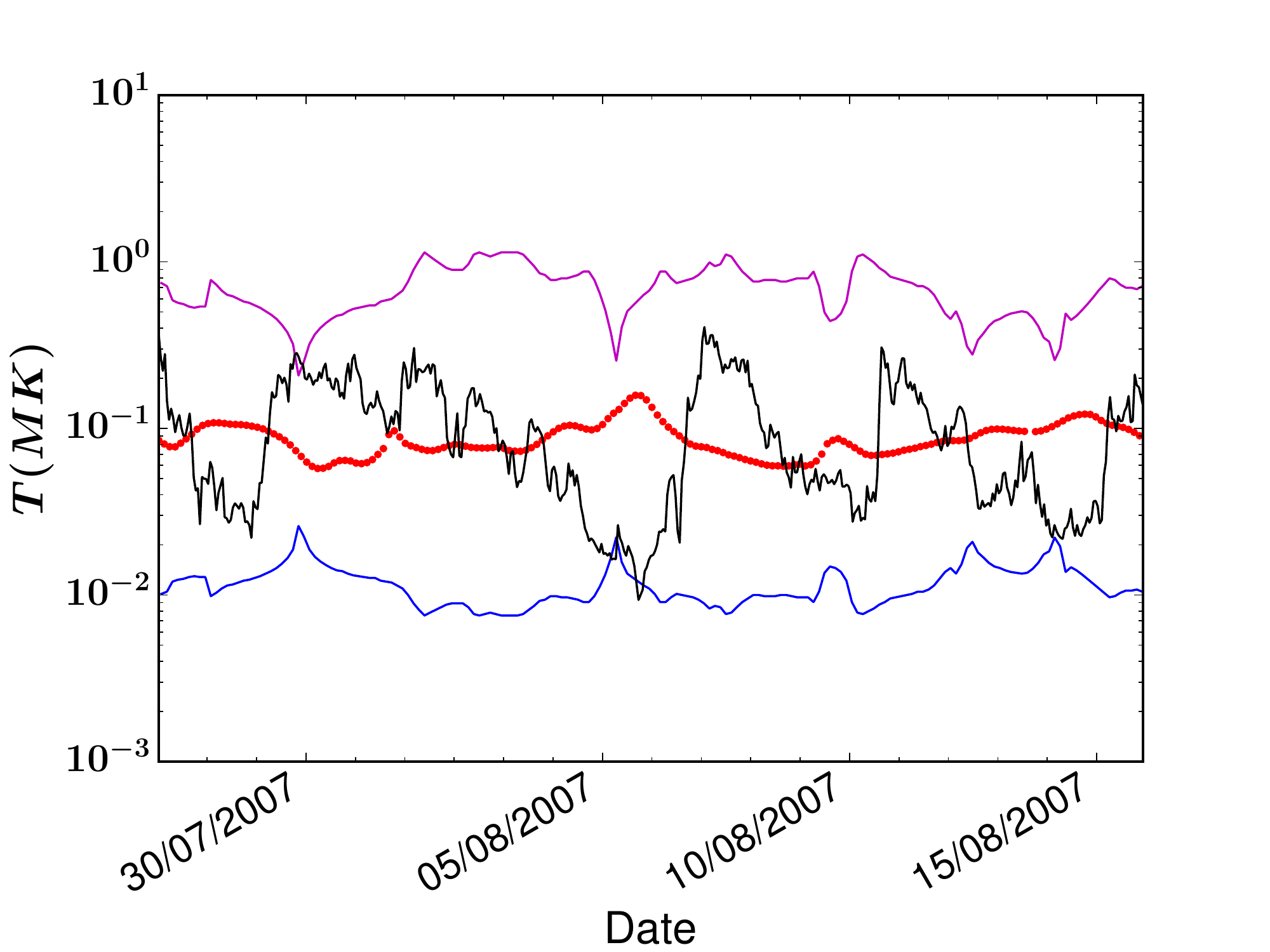}
\includegraphics[width=0.49\textwidth]{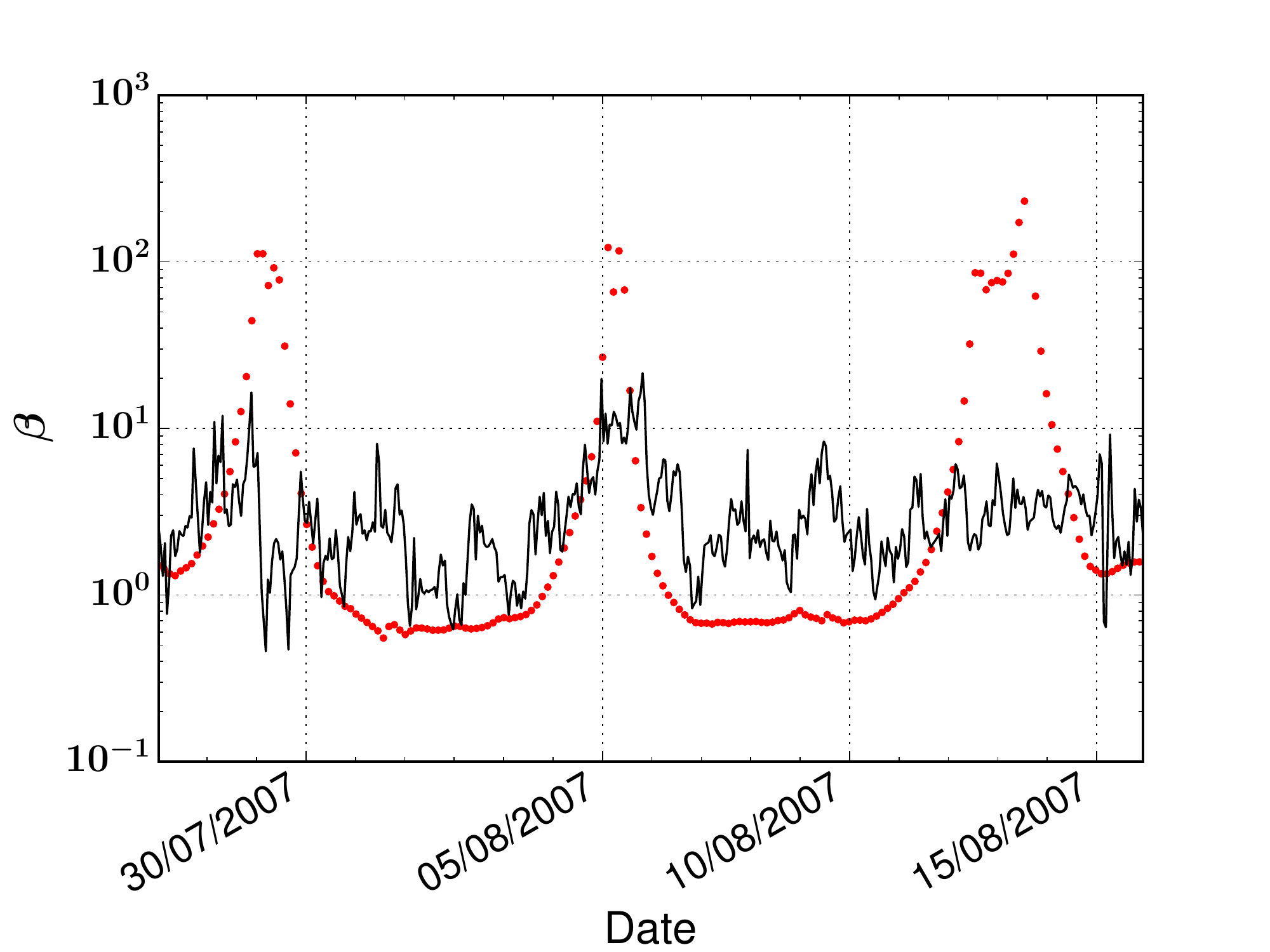}
\caption{Longitude cut-out as given by EUHFORIA and kinetic code at 1AU for i) the speed, ii) the number density, iii) the Alfv\'{e}nic Mach number, iv) the proton temperature of the solar wind and v) the plasma $\beta$ respectively, together with \textit{in-situ} observations. Blue corresponds to the kinetic model, red to the MHD model and black denotes the OMNI observations, whereas in the temperature panel magenta is the kinetic electron temperature curve and blue the kinetic proton temperature.}
\label{fig:all1au}
\end{center}
\end{figure}

In Figure \ref{fig:all1au}, the speed,
number density and temperature of the solar wind are shown for i) the MHD
model, ii) the kinetic model and iii) near-Earth observations using the OMNI dataset. From the speed plot
we conclude that both models reproduce the number of peaks, their position with respect to each other
and have a similar width. The most prominent difference between the
models and the observations appears
at the double peak of August 13th. EUHFORIA reproduces peaks of similar
amplitude as observed, but showing a higher global minimum at about 400
$\mathrm{km\ s^{-1}}$. The heights of the peaks and their relative ratio to one
another are not reproduced exactly by any model. The kinetic model
systematically overestimates the speeds varying from 400 to 800 $\mathrm{km\
s^{-1}}$. This can be explained by the fact that the acceleration in that model
is more efficient than the MHD case and continues at a significant rate at
distances larger than 0.1AU to about 50$\mathrm{R}_\odot$. This
indicates that the final velocity obtained with the kinetic model can be improved by
adapting the boundary conditions, for instance by lowering the empirical
solar wind speed close to the Sun. The second panel shows the number
density variations for the models and the observations. We observe that both
models and the OMNI observations of the number density agree roughly in order of
magnitude and in number of peaks, with the peaks not coinciding perfectly. The
peak amplitudes are about 2.5 times larger in the observations than in both
models, that are in agreement with one another. For the Alfv\'{e}nic Mach number
(third panel), we conclude that the Alfv\'{e}nic Mach number of the MHD
simulation is about a factor 3 higher than the average measured value most of the
time, with three lows that reach lower than the observed values. The average
observed proton temperature agrees with the EUHFORIA plasma temperature in order
of magnitude, varying from the kinetic proton temperature $T_p$ at its minimum
to the kinetic electron temperature $T_e$ at its maximum, while it is located
between the two species' temperatures, about an order of magnitude larger than
the kinetic proton temperature and an order of magnitude smaller than the
kinetic electron temperature, at all times. The variation profile of the
observations doesn't match with any model. Finally, the plasma $\beta$ shown in
the final panel for the MHD simulation lies most of the time at the lower limit
of the observed one, with three peaks that reach values of about a factor 3
higher than the observed highest values, showing the opposite trend when
compared to the Alfv\'{e}nic Mach number shown in the third panel. Thus we
conclude that, in the MHD simulation the magnetic field is higher with respect
to the plasma pressure and the number density than the one indicated by
observations at 1AU.

The same case corresponding to CR 2059 was analyzed in a comparative study
published in \citet{Gressl2014} for different observational inputs and MHD modeling
schemes. None of their models seemed to accurately represent observation. The case
directly comparable to ours was the one using GONG magnetograms and the WSA
empirical model and ENLIL for the MHD modeling (dark red curve of
lower panel in Figure 3). Before 27 July their maxima are not synchronous
nor do they reach the same values, but after that the model captures the times
of the peaks, but underestimates their value on the panel in the figure showing the speed.

%%%%%%%%%%%%%%%%%%%% \textit{Ulysses} 2007 %%%%%%%%%%%%%%%%%%%%%%%%%%%%%%%%

\begin{figure}[htbp]
\begin{center}
\includegraphics[width=0.49\textwidth]{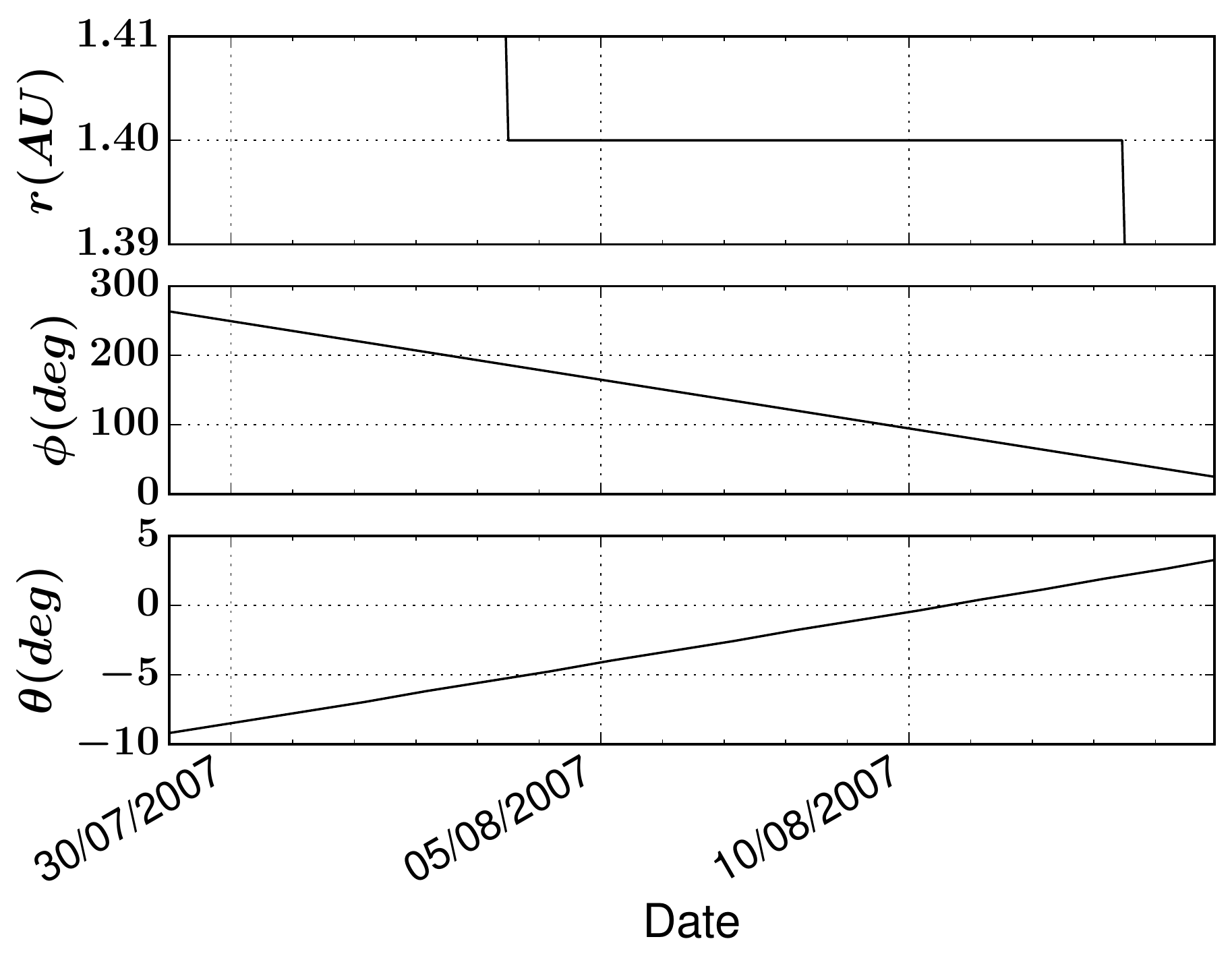}
\includegraphics[width=0.49\textwidth]{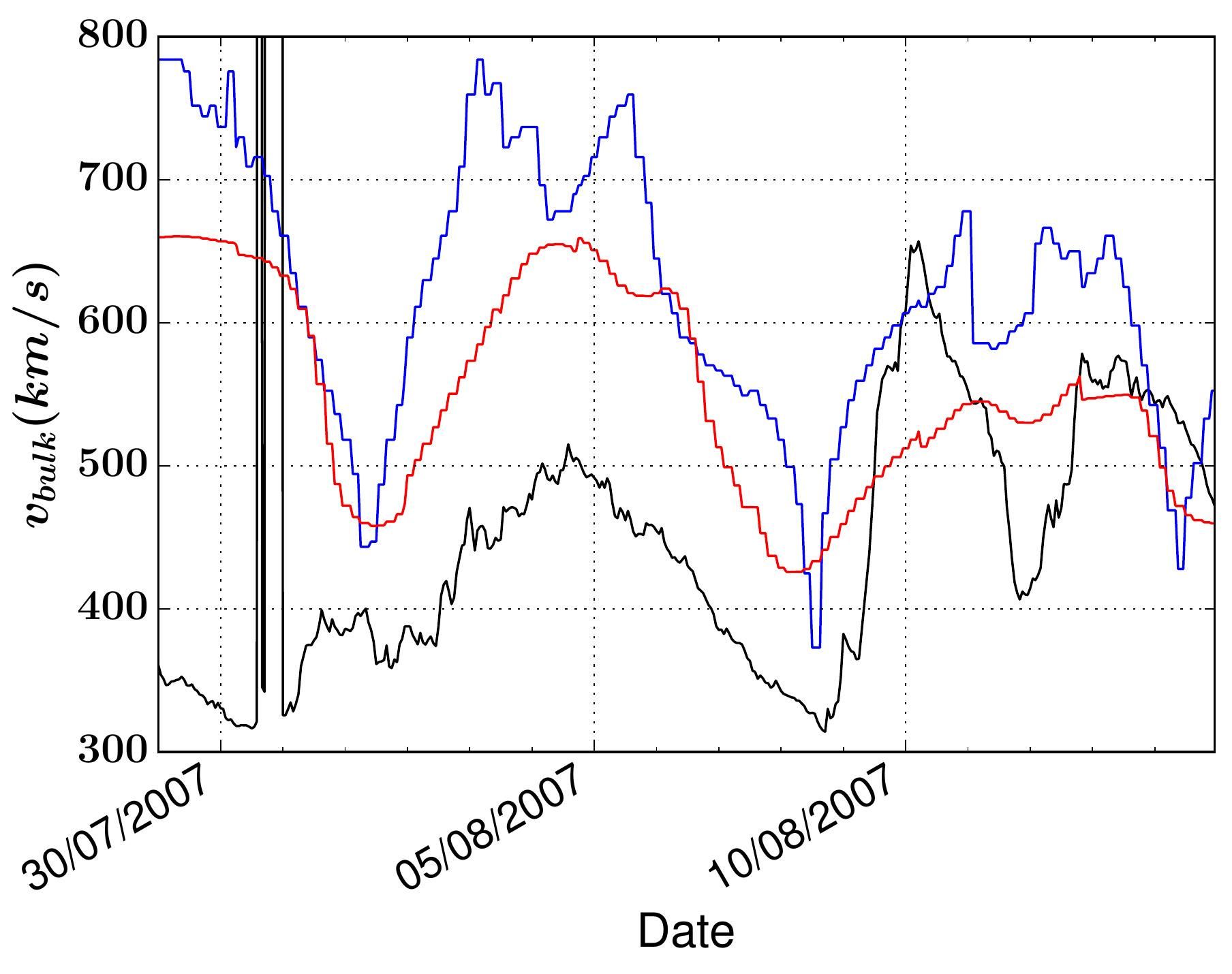}
\includegraphics[width=0.49\textwidth]{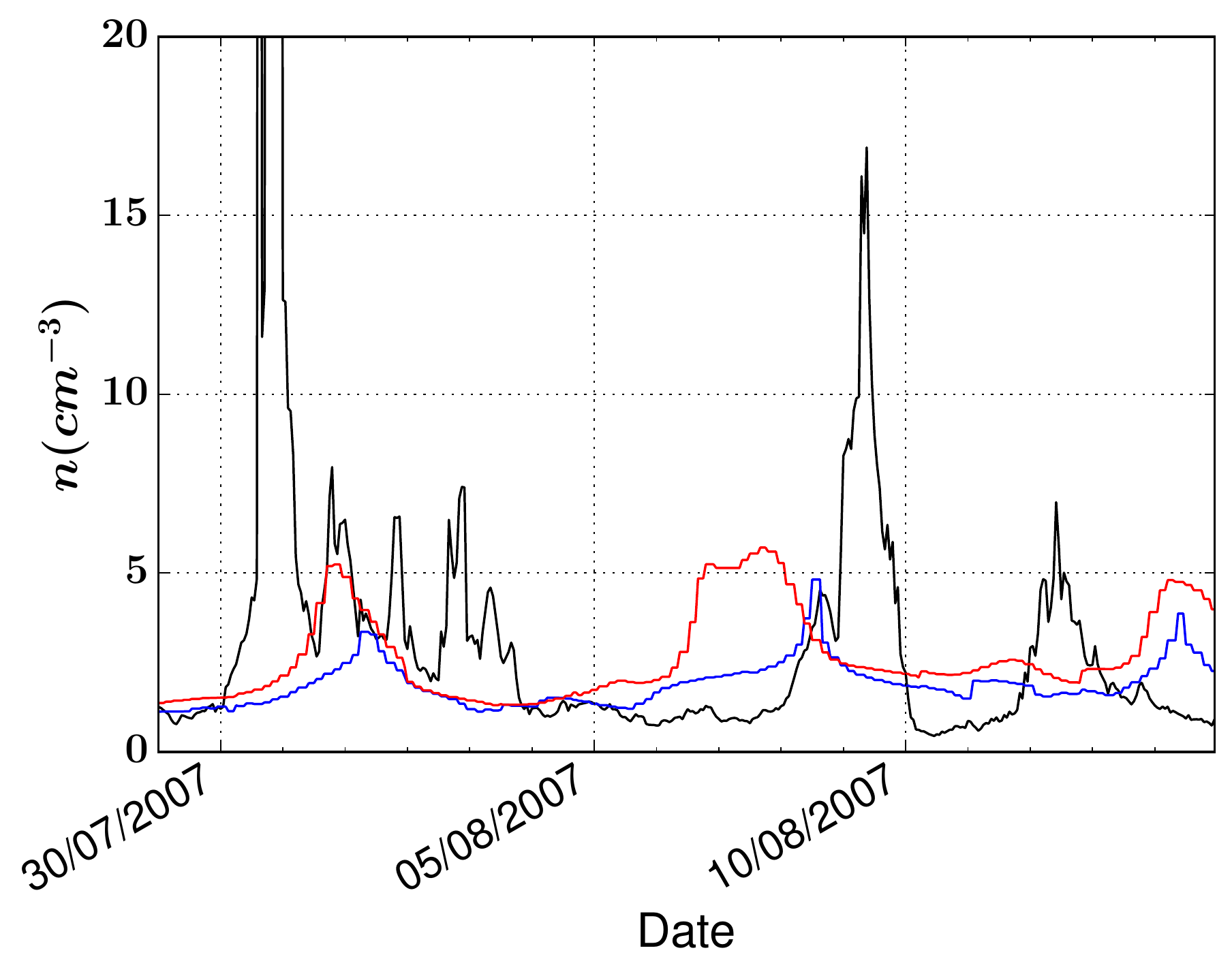}
\includegraphics[width=0.49\textwidth]{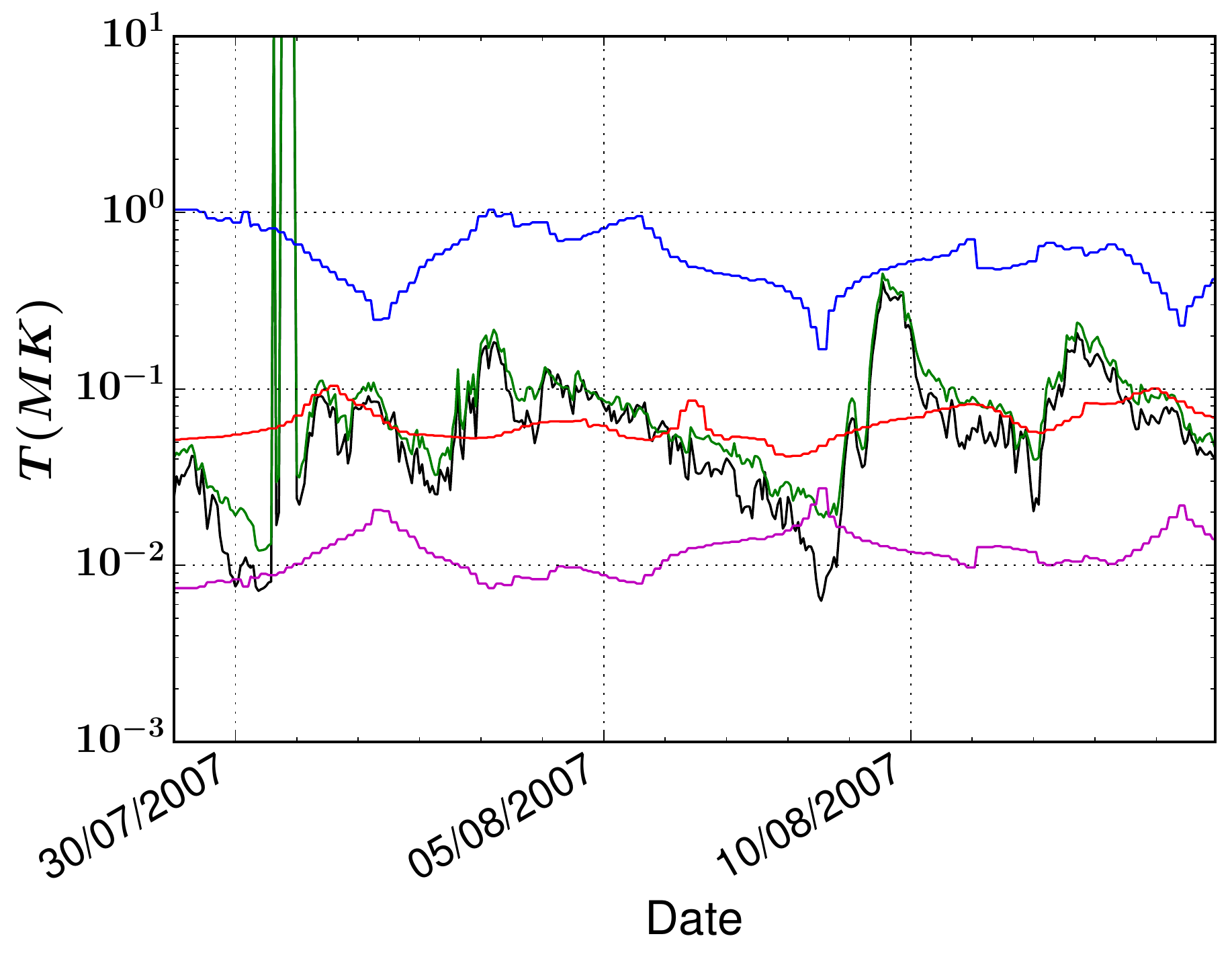}
\caption{Time variations as measured by \textit{Ulysses} (black) and calculated by the kinetic (blue) and the MHD (red) model, for i) \textit{top left panel}: the position of \textit{Ulysses} in heliographic coordinates, ii) \textit{top right panel}: the speed, iii) \textit{bottom left panel}: the number density and iv) \textit{bottom right panel}: the \textit{Ulysses} small (black) and large (green) proton, the kinetic proton (magenta), the kinetic electron (blue) and MHD (red) temperature.}
\label{fig:ul1}
\end{center}
\end{figure}

For a further comparison of the models with respect to observations, we overplot in
Figure \ref{fig:ul1} observations from \textit{Ulysses} at 1.4AU and at latitudes between
$-10^\circ$ and $5^\circ$ together with the results of the kinetic and MHD models. In
the first panel (top left), we show the trajectory of \textit{Ulysses} during the period
of interest. At the same 3D heliographic spherical coordinates we extract the
quantity of interest from the MHD and the kinetic model to accommodate further
comparisons. More specifically, for each $(r,\theta,\phi)$ position of \textit{Ulysses}
at each hour of a specific date, we choose the closest
available point for the specific resolution of each model. In the second panel,
we show the observed speed by \textit{Ulysses} with black, together with the kinetic
prediction in blue and the MHD results in red. All three exhibit
different time profiles
for each case. The kinetic model systematically overestimates the speeds
reaching maximum values of about 800 $\mathrm{km\ s}^{-1}$, whereas the MHD
model ranges from 400 to 650 $\mathrm{km\ s}^{-1}$, while at the same period the
\textit{Ulysses} measurements lie between 300 and 650 $\mathrm{km\ s}^{-1}$. The peaks
for both models are not well synchronized with the measured temporal velocity
profiles. For the density the two models don't reproduce the observed number of
peaks and they are not synchronized either. The number densities of the kinetic
and MHD models agree with each other varying from 1\,--\,5 $\mathrm{cm}^{-3}$ whereas the
observed number density profile is much more variable with a number of peaks and
reaching densities of 20 $\mathrm{cm}^{-3}$. The average observed proton
temperature\footnote{As shown in Figure \ref{fig:ul1}, there are two different
proton temperatures estimated, that in general bracket the real temperature at
1AU. As $T$-large we denote the integral in the 3D velocity space of the
distribution over all measured angles and energy bandwidths. The $T$-small is
calculated by the sum over all angles for a determined energy, then summing the
moments of the estimated spectrum of the plasma and by taking the radial
component of the temperature tensor
(\url{http://www.cosmos.esa.int/web/ulysses/swoops-ions-user-notes}).} agrees
with the plasma temperature predicted by the MHD model and lies in between the
kinetic proton (magenta) and kinetic electron (blue) temperatures of the kinetic
model, being one order of magnitude smaller than the electron and one order of
magnitude larger than the proton temperatures. No model reproduces in high accuracy the temporal variations of the observed temperature profile. The models show some correspondence with the
\textit{Ulysses} observations and the correct orders of magnitude are roughly reproduced,
but certainly there is need for improvement. To reach better
agreement with observations, we can modify the parameters in the semi-empirical
model of EUHFORIA to better adjust and reproduce the measurements in each case
individually.

\subsubsection{Heat Flux}

\citet{Pomoell2012} presented an analysis of the different ways that energy
source considerations can be used in MHD solar wind models. A generalized
formulation of the energy Equation \ref{eq:p} with an extra energy source term
at the right hand side $S$ was examined for different cases of
$S$. The relevant models studied therein include i) a model with a
polytropic index with spatial dependance $\Gamma(r)$ and ii) a model with a
polytropic index fixed at $\gamma=5/3$ constant in the entire coronal volume.
In particular, the authors showed that a steady-state solar wind solution
accelerated by a given non-adiabatic polytropic wind can equivalently be
re-written using an energy source term, given by:
\begin{equation}
S=\nabla\cdot \Bigg[\mathbfit{v}P\bigg( \frac{1}{\Gamma-1}-\frac{1}{\gamma-1}\bigg) \Bigg] {,}
\end{equation}
with $\mathbfit{v}$, $P$ and $\Gamma$ being obtained by the model, as explained in \citet{Pomoell2012}, with the non-adiabatic index $\Gamma(r)=1.5$ for our case.

\begin{figure}[htbp]
\begin{center}
\includegraphics[width=0.8\textwidth]{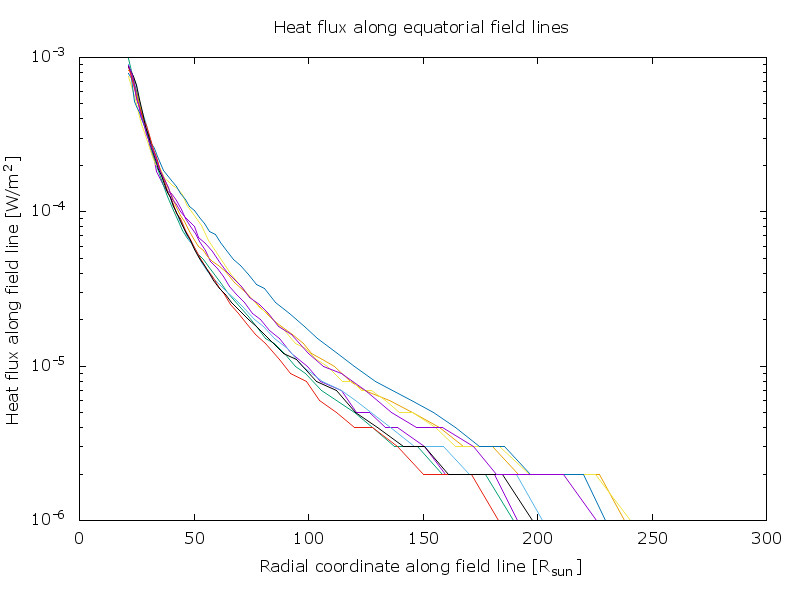}
\caption{Heat flux along magnetic field lines in the equatorial plane as given by EUHFORIA.}
\label{fig:heatfluxeuf}
\end{center}
\end{figure}

In this section, we will discuss the heat flux of the MHD and the kinetic models.
%trying to dive more into the physical mechanisms driving the two models and
%holding a first comparison on their main differences and the places that they
%arise in the coronal volume from close to large heliocentric distances.
According to the formulation presented above we have that the analytic
expression of the energy flux responsible for accelerating the wind in the MHD model is given by $\mathbfit{v}P\bigg(
\frac{1}{\Gamma-1}-\frac{1}{\gamma-1}\bigg)$. In Figure \ref{fig:heatfluxeuf},
we present the heat flux of the MHD model along selected magnetic field lines
as a function of distance from
0.1AU to 2AU in the equatorial plane.

\begin{figure}[htbp]
\begin{center}
\includegraphics[width=\textwidth]{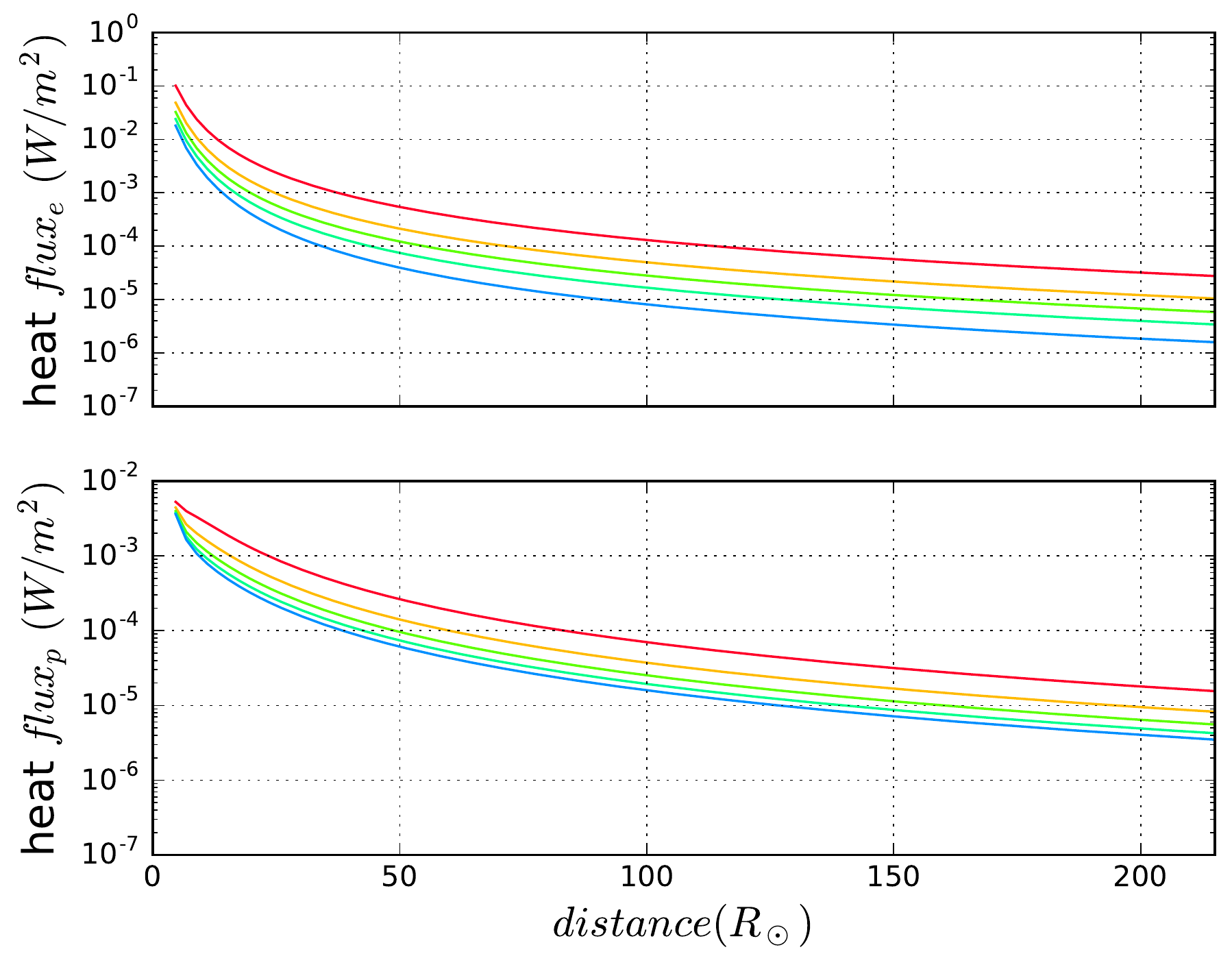}
\caption{Electron (top panel) and proton (bottom panel) heat fluxes for CR2059 with red, orange, light green, green and blue curves corresponding to $\kappa$ values of 3, 4, 5, 6 and 7, respectively.}
\label{fig:heatflux}
\end{center}
\end{figure}

In Figure \ref{fig:heatflux}, we illustrate the radial
profiles of the electron (top panel) and proton (bottom panel) heat fluxes, as
calculated by the kinetic model for an exobase at 2.5$\mathrm{R}_\odot$, 1MK
electron and proton temperatures at the exobase $T_e=T_p=1\mathrm{MK}$ and the same
densities as our default case for $\kappa$-indexes 3, 4, 5, 6 and 7
corresponding to red, orange, light green, green and blue respectively. We
observe that the more suprathermal particles are present, \textit{i.e.}\ lower $\kappa$, the higher the
heat flux is. The electron heat flux is higher by an order of magnitude than the
proton heat flux close to the Sun and it decreases faster than the proton heat
flux with the heliocentric distance to reach similar values at 1AU. 
We conclude that the differences between the proton heat flux curves corresponding to different $\kappa$-indexes are smaller than the respective electron flux curves.
In general, the heat fluxes are overestimated by exospheric models, since the corresponding VDFs have the highest possible anisotropy due to lack of collisions. As shown in \cite{Zouganelis05}, exospheric models and kinetic simulations that include collisions are in good agreement, making the exospheric model a convenient tool for the study of weakly collisional plasmas. Both the exospheric model and the collisional simulations found heat fluxes several times larger than the classical value, suggesting that the classical formulation is not appropriate for weakly collisional plasmas, since it was based on the assumption of a collision-dominated medium.
If one really needs to prescribe a realistic heat flux, then collisions need to be taken into account and further improvements in the kinetic model need to be considered. The inclusion of interactions, through \textit{e.g.}\
Alfv\'{e}n or whistler waves, will decrease the anisotropies and it will improve
the higher moments, \textit{i.e.}\ temperatures and heat fluxes for the considered
species~\citep{Pierrard2011Schlickeiser,Pierrard14,Voitenko2015}. Using such sophisticated schemes to improve the heat flux agreement with observations is outside of the scope of this paper, due to i) the consequent increased computational expense, ii) the fact that the temperatures are not the most important geo-effective parameters, making these improvements rather impractical for future operational space weather applications. Thus, the heat flux profiles for the electron and protons, that are quantified by the kinetic
model (Figure \ref{fig:heatflux}) can serve as upper limits and are more physics-based than the MHD heating prescriptions that are based on the empirical
determination of the polytropic index value.

The heat flux profiles along the magnetic field lines of the MHD model are in
agreement in order of magnitude with the profiles of the protons of the kinetic
model closer to the Sun, but they drop faster further out reaching electron heat
flux values at the orbit of the Earth. Our results for the heat flux for the MHD
model depicted in Figure
\ref{fig:heatfluxeuf} are roughly in agreement with the
literature~\citep{Hellinger2013JGR,Stverak2015}. In accordance to the kinetic
model results, the electron heat flux is higher than the proton heat flux
according to \cite{Hellinger2013JGR} and \cite{Stverak2015}, respectively, with
our results roughly being closer to the electron energetics profile, but at 1AU
approaching the heat flux values expected for the protons.

%%%%%%%%%%%%%%%%%%%%%%%%%%%%%%%%%%%%%%%%%%%%%%%%%
%%%%%%%%%%%%%%%%%%%%%%%%%%%%%%%%%%%%%%%%%%%%%%%%%
%%%%%%%%%%%%%%%%%%%%%%%%%%%%%%%%%%%%%%%%%%%%%%%%%
%%%%%%%%%%%%%%%%%%%%%%%%%%%%%%%%%%%%%%%%%%%%%%%%%
%\section{Comparisons}

\section{Discussion}

In this study, we are interested in making a first comparison between single
fluid and kinetic models, that have the potential to be used in space weather
applications. The two models are very different in nature, making use of very
different formulations for the plasma physics. EUHFORIA is a single fluid code
that uses the MHD equations to describe the plasma. Whereas the kinetic model used
in this study begins by making an observationally-inspired assumption for the
VDFs of the electron and proton species, that are considered to constitute the
solar wind plasma, and based on that all the physically interesting quantities
are calculated as moments over the velocity space. EUHFORIA is
observationally-driven and provides three-dimensional information that
accounts for stream interactions and complicated magnetic topologies. The
kinetic model is a semi-analytic model solving for the plasma characteristics
along a magnetic field line and accounting for heating and acceleration in a
self consistent way. Note that the electric field (used in the kinetic model) is
hidden in the pressure term \cite[as shown by][]{Parker2010}. Parker
explained that the momentum equation of the electrons is given by
\begin{equation}
\frac{\mathrm{d}p_e}{\mathrm{d}r}+neE_\mathrm{LS}=0 {,}
\end{equation}
which is the hydrodynamic condition, with $E_\mathrm{LS}$ the Lemaire-Scherer electric field ~\citep{Lemaire}.
In \cite{Parker2010}, Parker showed that the electric field of Lemaire-Scherer is in the fluid approach given by
\begin{equation}
E_\mathrm{LS}=-\frac{1}{ne}\frac{\mathrm{d}p_e}{\mathrm{d}r}=\frac{m_p}{e}\left( v\frac{\mathrm{d}v}{\mathrm{d}r}+\frac{GM_\odot}{r^2}\right) {,}
\end{equation}
which finally, taking into account that the Pannekoek-Rosseland electric field is $E_\mathrm{PR}=\frac{m_pg}{e}$, is written as
\begin{equation}
E_\mathrm{LS}=E_\mathrm{PR}+\frac{m_pv}{e} \frac{\mathrm{d}v}{\mathrm{d}r} {.}
\end{equation}
The Lemaire-Scherer electric field (used in exospheric models) is several times
larger than the Pannekoek-Rosseland one, corresponding to hydrodynamic
equilibrium that is used to describe the solar wind expansion. It is able to
lift and accelerate to supersonic speeds the initially slow and heavy protons
through trapping the fast and light electrons, while keeping the
quasi-neutrality and almost zero-current condition. Any difference in the bulk
speeds of the two populations would lead to the generation of a current, which
due to Amp\`{e}re's law has to remain small.

From Figure~\ref{fig:eufvrbr3}, we deduce that at large radial distances
from the Sun, the MHD current sheet gets expanded and becomes thicker, evolving
from about $2^\circ$ at the inner boundary to about $10^\circ$ at 1AU, while the fine
structures that appear in the 0.1AU longitudinal and latitudinal map get
diffused and smoothed. On the contrary, in Figure~\ref{fig:kinslice} we see that
the fine structures and the current sheet size don't change under the kinetic
approach, as there are no stream interactions taken into account. Both models
give speeds of the same order of magnitude at every altitude and have most of
their acceleration taking place already before 1AU. The kinetic model shows a
more efficient acceleration, reaching terminal speeds of about 100 $\mathrm{km\
s^{-1}}$ larger than the MHD one up to 1AU. The acceleration in the MHD approach
is taking place due to the reduced polytropic index,
while in the kinetic approach the acceleration is related to the
induced electric field that assures quasi-neutrality and equal outward electron
and proton fluxes. Regarding the number density, both models give densities of
the same order of magnitude, with the kinetic model though showing a sharper
profile that decreases faster outwards. The number density of the kinetic model
is 20\% and 30\% lower at 1AU and 1.4 AU respectively than the MHD number
density. Moreover, for the temperature in the MHD approach, as is shown in the
bottom row of Figure \ref{fig:eufvrbr3}, there is a reversal of the hot-cold
regions from the inner boundary up to large distances. A faster cooling takes
place at higher latitudes, making the initially colder equatorial region appear
hotter at large radial distances with respect to the rest of the latitudes.
Furthermore, the temperature ranges fall by a factor 4 from 0.1AU up to 1AU,
and continue decreasing by 30\% up to the orbit of \textit{Ulysses}. For the kinetic
approach, the proton temperature only falls by factor two up to the Earth's
orbit and seems constant up to 1.4 AU, while the electron temperature drops by a
factor 5 up to 1AU and doesn't seem to vary much from there on. Contrary to the
MHD single fluid results, in the kinetic case there is no temperature profile
change and the cooling seems to take place uniformly at all latitudes. There is
a structure close to the equator at a heliographic longitude of $-70^\circ$ that
appears like a spike in contact with the current sheet in both Figures \ref{fig:eufvrbr3} and \ref{fig:kinslice}. In Figure \ref{fig:eufvrbr3}
though, in the middle and right panels, corresponding to large distances, that
spike appears to change inclination from pointing to the left of the figure at
the panels corresponding to 0.1AU to pointing to the right from 1AU outwards,
unlike Figure \ref{fig:kinslice}, where the spike's inclination remains
unchanged. This difference is then likely due to stream interactions that are
captured in the 3D MHD model, while these are excluded in the
essentially 1D, radial field kinetic approach.

\section{Conclusions}

After the parallelization of the kinetic model and its use in a quasi-3D
approach we linked it to more robust 3D MHD models and compared both to
observations at 1AU, using similar boundary conditions at
$21.5\mathrm{R}_\odot$. When the two models were compared, starting from the
same boundary conditions, the kinetic model gives systematically higher speeds
than the MHD model at large radial distances. This is due to the fact that the
acceleration of the solar wind continues at a higher rate in the kinetic model
after $21.5\mathrm{R}_\odot$. The acceleration mechanism in the kinetic model is
due to the induced electric field that assures quasi-neutrality and prevents
charge separation and also "bounds" the two species to move with the same bulk
speed. There is no explicit heating term in the MHD equations used by EUHFORIA.
The MHD model accelerates further the solar wind due to the reduced polytropic
index $\gamma$, but at a very slow rate accounting for an acceleration of about
50 $\mathrm{km\ s}^{-1}$ from 0.1AU to 1AU.

The exospheric models overestimate the heat flux, especially for the electrons
that have a thermal speed comparable to their bulk speed, but such a heat flux
can be improved by inclusion of interactions through waves, \textit{e.g.}\ Alfv\'{e}n or
whistler waves~\citep{Pierrard2011Schlickeiser,Pierrard14,Voitenko2015}. The
heat flux calculated by the kinetic exospheric model can be used as an upper
limit for more physically-driven heat flux prescriptions in a global MHD model.
The heat flux of the kinetic model is in qualitative agreement with other studies \citep{Hellinger2013JGR,Stverak2015} and the heat flux profile of the MHD models close to the Sun resembles the proton profile, only to decrease faster outwards resembling the electron profile at 1AU.

In the exospheric model the fast electrons are slowed down and the protons are
accelerated by the Lemaire-Scherer electric field that eventually leads to the
observed supersonic solar wind. As shown by \cite{Parker2010} the acceleration
of the solar wind in collisionless plasmas to supersonic values lies in the
hydrodynamic equation in combination to the mass ratio between electrons and
protons. According to Parker, the exospheric model describes a very efficient
heat transport mechanism with an electron temperature that decreases very slowly
at large distances and through the induced electric field it elevates the
protons and causes the transonic solar wind.

There is some agreement in the high- and slow-speed streams in the velocity
profiles for both OMNI and \textit{Ulysses} observations. The high speed positions of the
models are better synchronized for OMNI rather than \textit{Ulysses} observations, as we
showed earlier. The number densities of both models were approximately in the
same order of magnitude with the observed ones at 1AU and at the orbit of
\textit{Ulysses}. The MHD temperature is one order of magnitude smaller than the electron
temperature $T_e$ and one order of magnitude larger than the proton temperature
$T_p$ of the kinetic model. The average observed proton temperatures agreed with
the temperature predicted by the MHD model in order of magnitude and thus were
also lying in the range between the electron and proton kinetic temperatures.

In this study we assumed that the $\kappa$-value is independent of the radial
distance, but observations such as~\cite{Maksimovic05} indicate that the
$\kappa$ can actually change as the distance from the Sun increases. However, the fitting model
used in~\cite{Maksimovic05} was not with a Kappa distribution for the full
range, but a sum of a Maxwellian for the core and a Kappa function for the halo,
so that the parameter $\kappa$ does not represent the same quantity as in the
model used in the current study. The density ratio between the core and the halo
remains constant with the distance, as analyzed by~\cite{Pierrard16}. A more realistic exobase profile with temperatures having a latitudinal and even longitudinal variation can be taken into account and will be the next step towards a fully 3D kinetic numerical code of the solar corona and the solar wind.

The kinetic model is a semi-analytic model ignoring stream interactions and thus conserves the slow and fast wind distributions for every radial distance. 
Accounting for stream interactions as the solar wind propagates outwards will be one of the main points of interest for a more realistic 3D model. 
Shocks associated to sharp velocity gradients can be included in an empirical way or by using more sophisticated kinetic models including collisions and wave-particle interactions \citep[see \textit{e.g.}][]{VivianeLazar_book}.
Here we have ignored the spiral shape of the magnetic field in the calculation of the moments in the kinetic model, adopting purely radial magnetic fields, because as argued in~\cite{Pierrard01meyer} this aspect won't affect the main average quantities apart from the temperature anisotropies. 
But in the future we are planing on including the spiral magnetic field effects in a similar study. 
\cite{Pierrard01meyer} quantified the effect that the spiral magnetic field topology has on the particle temperatures and their anisotropies. 
More specifically, in the radial case the electron temperature is slightly underestimated, whereas the opposite is true for the proton species. 
The temperature anisotropies are overestimated by the kinetic model in comparison to observations \citep{Lemaire01}. 
Another important aspect that can be improved and would provide a deeper comparison between operational MHD codes and kinetic exospheric ones, would be to change the formulation of the kinetic model, so that boundary conditions for the speed, the density and temperature directly from MHD models can be used at each grid point including the magnetic field information. 
These aspects will allow us to reach more fundamental conclusions about the two different models and will upgrade the kinetic exospheric model into a computational equivalent to the robust 3D MHD code. 
When the 3D magnetic field topology from the source surface, through the Schatten current sheet region, and throughout the region covered by the MHD model, is used directly within the kinetic description, we can use its predicted heat fluxes and higher order moment information to turn the model into a self-consistent hybrid kinetic-MHD modeling tool.

%%%%%%%%%%%%%%%%%%%%%%%%%%%%%%%%%%%%%%%%%%%%%%%%%%%%%%%%%%%%%%
%%%%%%%%%%%%%%%%%%%%%%%%%%%%%%%%%%%%%%%%%%%%%%%%%%%%%%%%%%%%%%

%%%%%%%%%%%%%%%%%%%%%%%%%%%%%%%%%%%%%%%%%%%%%%%%%%%%%%%%%%%%%%
%%%%%%%%%%%%%%%%%%%%%% BIBLIOGRAPHY %%%%%%%%%%%%%%%%%%%%%%%%%%%%%%%

%\bibliographystyle{abbrv}

%%%%%%%%%%%%%%%%%%%%%%%%%%%%%%%%%%%%%%%%%%%%%%%%%%%
%% Authors Names
%
% \author[addressref={},corref,email={}]{\inits{}\fnm{}\lnm{}\orcid{}}

%%%%%%%%%%%%%%%%%%%%%%%%%%%%%%%%%%%%%%%%%%%%%%%%%%%
%% Runningheads
%
%\runningauthor{}
%\runningtitle{}

%%%%%%%%%%%%%%%%%%%%%%%%%%%%%%%%%%%%%%%%%%%%%%%%%%%
%% Affilations
%% id shold be the same with \author addressref value.
%\address[id={}]{}

%%%%%%%%%%%%%%%%%%%%%%%%%%%%%%%%%%%%%%%%%%%%%%%%%%%
%%% Abstract
%\begin{abstract}
%\end{abstract}

%%%%%%%%%%%%%%%%%%%%%%%%%%%%%%%%%%%%%%%%%%%%%%%%%%%
%% Keywords
%
%\keywords{}

%-------------------------------------------------

%%%%%%%%%%%%%%%%%%%%%%%%%%%%%%%%%%%%%%%%%%%%%%%%%%%
%% Sections
%
% \section{}%\label{s:?}

%% Figure
%
% \begin{figure}
% \centerline{\includegraphics[width=0.5\textwidth,clip=]{<fig.eps>}}
% \caption{}%\label{fig:?}
% \end{figure}

%% Table
%
% \begin{table}
% \caption{}%\label{tbl:?}
% \begin{tabular}{}
% \hline
% \multicolumn{2}{c}{<>}
% <data>
% \hline
% \end{tabular}
% \end{table}

%%%%%%%%%%%%%%%%%%%%%%%%%%%%%%%%%%%%%%%%%%%%%%%%%%%%%%%%%%%%%%%%%%%%%%%%%%%
%% Appendix
%
% \appendix

%%%%%%%%%%%%%%%%%%%%%%%%%%%%%%%%%%%%%%%%%%%%%%%%%%%%%%%%%%%%%%%%%%%%%%%%%%%
% Acknowledgements
%
 \begin{acks}
SPM acknowledges financial support by the FWO and NASA Living with a Star grant number NNX16AC11G. This research was supported by projects GOA/2015\,--\,014 (KU Leuven, 2014\,--\,2018), and the Interuniversity Attraction Poles Programme initiated by the Belgian Science Policy Office (IAP P7/08 CHARM). 
 \end{acks}

{\footnotesize\paragraph*{Conflict of Interest} The authors declare that they have no conflict of interest.}

%%% %%%%%%%%%%%%%%%%%%%%%%%%%%%%%%%%%%%%%%%%%%%%%%%%%%%%%%%%%%%
%% Bibliography
%
% Using BibTeX
%
% \bibliographystyle{spr-mp-sola}
% \bibliography{<bib file>}
\bibliographystyle{spr-mp-sola}
\bibliography{allpapers}
%
% Without BibTeX
% \begin{thebibliography}{}
% \bibitem[\protect\citeauthoryear{Author}{Year}]{key}
%   <bibliographical entry>
%
% \bibitem[\protect\citeauthoryear{}{}]{}
%
% \end{thebibliography}

\end{article}
\end{document}